\RequirePackage{amsmath}
\documentclass[conference]{IEEEtran}

\usepackage[T1]{fontenc}

\usepackage{multirow}
\usepackage{graphicx}
\usepackage{booktabs}
\usepackage{subcaption}
\usepackage{caption}
\usepackage{tabularx}
\usepackage{etoolbox}
\usepackage{cite}
\usepackage{amsmath,amssymb,amsfonts}
\usepackage{algorithmic}
\usepackage{textcomp}

\usepackage{csquotes} %

\usepackage[hidelinks]{hyperref}

\usepackage{stmaryrd}
\usepackage{xcolor}

\usepackage{microtype}

\makeatletter
\renewcommand{\sectionautorefname}{\S\@gobble}
\renewcommand{\subsectionautorefname}{\S\@gobble}
\renewcommand{\subsubsectionautorefname}{\S\@gobble}

\renewcommand{\paragraphautorefname}{\S\@gobble}
\makeatother

\AtBeginEnvironment{equation}{\small}

\usepackage{acro}[=v2]

\DeclareAcronym{defi}{
    short = DeFi,
    long = Decentralized Finance
}

\DeclareAcronym{amm}{
    short = AMM,
    long = Automated Market Maker
}

\DeclareAcronym{amms}{
    short = AMMs,
    long = Automated Market Makers
}

\DeclareAcronym{cfmm}{
    short = CFMM,
    long = Constant Function Market Maker
}

\DeclareAcronym{xrpl}{
    short = XRPL,
    long = XRP Ledger
}

\DeclareAcronym{dao}{
    short = DAO,
    long = Decentralized Autonomous Organization
}

\DeclareAcronym{dex}{
    short = DEX,
    long = Decentralized Exchange
}

\DeclareAcronym{dexs}{
    short = DEXs,
    long = Decentralized Exchanges
}

\DeclareAcronym{dapps}{
    short = DApps,
    long = Decentralized Applications
}

\DeclareAcronym{dapp}{
    short = DApp,
    long = Decentralized Application
}

\DeclareAcronym{cam}{
    short = CAM,
    long = Continuous Auction Mechanism
}

\DeclareAcronym{gbm}{
    short = GBM,
    long = Geometric Brownian Motion
}

\DeclareAcronym{g3m}{
    short = G3M,
    long = Geometric Mean Market Maker
}

\DeclareAcronym{cpmm}{
    short = CPMM,
    long = Constant Product Market Maker
}

\DeclareAcronym{cpmms}{
    short = CPMMs,
    long = Constant Product Market Makers
}

\DeclareAcronym{g-amm}{
    short = G-AMM-DEX,
    long = Generic AMM-based DEX
}

\DeclareAcronym{mev}{
    short = MEV,
    long = Maximal Extractable Value
}

\DeclareAcronym{ctor}{
    short = CTOR,
    long = Canonical Transaction Ordering
}

\DeclareAcronym{clob}{
    short = CLOB,
    long = Central Limit Order Book
}

\DeclareAcronym{clobs}{
    short = CLOBs,
    long = Central Limit Order Books
}

\DeclareAcronym{cexs}{
    short = CEXs,
    long = Centralized Exchanges
}

\DeclareAcronym{cex}{
    short = CEX,
    long = Centralized Exchange
}

\DeclareAcronym{lps}{
    short = LPs,
    long = Liquidity Providers
}

\DeclareAcronym{lvr}{
    short = LVR,
    long = Loss-Versus-Rebalancing
}

\DeclareAcronym{cdf}{
    short = CDF,
    long = Cumulative Distribution Function
}

\DeclareAcronym{cdfs}{
    short = CDFs,
    long = Cumulative Distribution Functions
}

\DeclareAcronym{tvl}{
    short = TVL,
    long = Total Value Locked
}

\DeclareAcronym{lptokens}{
    short = LP Tokens,
    long = Liquidity Provider Tokens
}

\DeclareAcronym{evm}{
    short = EVM,
    long = Ethereum Virtual Machine
}

\DeclareAcronym{spe}{
    short = SPE,
    long = Spot-Exchange Rate
}

\DeclareAcronym{sp}{
    short = SP,
    long = Spot-Price
}

\DeclareAcronym{ep}{
    short = EP,
    long = Effective Price
}

\DeclareAcronym{usdc}{
    short = USDC,
    long = USD Coin
}

\DeclareAcronym{lobs}{
    short = LOBs,
    long = Limited Order Books
}

\DeclareAcronym{ukcbt}{
    short = UK CBT,
    long = UK Centre for Blockchain Technologies
}

\DeclareAcronym{ubri}{
    short = UBRI,
    long = University Blockchain Research Initiative
}

\begin{document}

\title{AMM-based DEX on the XRP Ledger}

\IEEEoverridecommandlockouts
\author{
    \IEEEauthorblockN{Walter Hernandez Cruz$^{1,3,4}$, Firas Dahi$^{1}$, Yebo Feng$^{\star,2}$, Jiahua Xu$^{1,3,4}$, Aanchal Malhotra$^{5}$, Paolo Tasca$^{1,3,4}$}
    \IEEEauthorblockA{$^{1}$Centre for Blockchain Technologies, University College London, $^{2}$Nanyang Technological University, \\ $^{3}$Exponential Science, $^{4}$UK Centre for Blockchain Technologies,$^{5}$Ripple Labs Inc.}
    \thanks{$^{\star}$Corresponding author: Yebo Feng (yebo.feng@ntu.edu.sg).}
}

\maketitle

\begin{abstract}

\ac{amm}-based \ac{dexs} are crucial in \ac{defi}, but Ethereum implementations suffer from high transaction costs and price synchronization challenges. To address these limitations, we compare the \ac{xrpl}-\ac{amm}-\ac{dex}, a protocol-level implementation, against a \ac{g-amm} on Ethereum, akin to Uniswap's V2 \ac{amm} implementation, through agent-based simulations using real market data and multiple volatility scenarios generated via \ac{gbm}. Results demonstrate that the \ac{xrpl}-\ac{amm}-\ac{dex} achieves superior price synchronization, reduced slippage, and improved returns due to \ac{xrpl}'s lower fees and shorter block times, with benefits amplifying during market volatility. The integrated \ac{cam} further mitigates impermanent loss by redistributing arbitrage value to \ac{lps}. To the best of our knowledge, this study represents the first comparative analysis between protocol-level and smart contract \ac{amm}-based \ac{dex} implementations and the first agent-based simulation validating theoretical auction mechanisms for \ac{amm}-based \ac{dexs}.

\end{abstract}

\begin{IEEEkeywords}
Automated Market Maker, XRP Ledger, Decentralized Finance, Continuous Auction Mechanism
\end{IEEEkeywords}

\section{Introduction}
\label{sec:intro}

\acl{defi} (\acs{defi}) has transformed financial services by using blockchain technology to offer new, transparent financial services without traditional intermediaries \cite{Zetzsche2020DecentralizedFinance,Xu2022b,Xu2022e,Luo2025PiercingReappraised,Ibanez2020c,Ibanez2025Triple-EntryKin,Liu2025DynaShard:Management} like banks \cite{Xu2022d}, lending platforms \cite{Arora2024SecPLF:Attacks,Perez2020liquidations,Xu2025Auto.gov:DeFi}, centralized exchanges \cite{xu2021dexAmm,Xu2021b,Dubovitskaya2021b}, insurance companies \cite{Cousaert2022,Braun2019c}, and wealth managers \cite{Xu2022g,Cousaert2021}. A key part of \ac{defi} are \ac{dexs} powered by \ac{amms}, first introduced by Bancor in 2017 and made popular by Uniswap \cite{Hertzog2017BancorTokens,uniswap_wp}. These \ac{dexs} use smart contracts and algorithms to enable trading without traditional market makers. The most common type of \ac{amm} used by \ac{dexs} is the \ac{cfmm}, with Uniswap V2's \ac{cpmm} being the most widely used \cite{Adams2020,uniswap_wp}.

Most \ac{amm}-based \ac{dexs} run on Ethereum and face several problems: high fees, large price changes during trades (slippage), impermanent loss for liquidity providers, and outdated prices compared to other markets \cite{most_AMMs_on_Ethereum,xu2021dexAmm,Fritsch2024MeasuringLiquidity}. These issues originate from \ac{amm}-based \ac{dexs}' design and their underlying infrastructure, which can lead to losses when off-chain prices move \cite{Milionis2022QuantifyingMakers,Milionis2022AutomatedLoss-Versus-Rebalancing,Fritsch2024MeasuringLiquidity}. Because \ac{amm}-based \ac{dexs} often quote outdated prices compared to real-time \ac{cexs}, arbitrageurs can profit from these differences, which usually results in impermanent losses for \ac{lps}, as the opportunity cost of providing liquidity often outweighs the fees earned \cite{Fritsch2024MeasuringLiquidity,Milionis2022AutomatedLoss-Versus-Rebalancing,Milionis2022QuantifyingMakers}, particularly in Uniswap V3 \cite{Cartea2022DecentralisedProvision,Cohen2024InefficiencySimulations,Sabate-Vidales2023TheSimulation}. 
While profiting from price discrepancies, arbitrageurs face slippage losses when the effective trade price differs from the initially quoted price. This occurs because \ac{amm}-based \ac{dexs} prices do not immediately update to reflect external market changes or new transactions between trade submission and finalization.

Learning from these issues, the \ac{xrpl}-\ac{amm}-\ac{dex} \cite{github} presents an alternative to existing \ac{amm}-based \ac{dexs}. It works on the \acl{xrpl} and aims to reduce price slippage during trades, keep prices in line with other external off-chain markets, and work more efficiently. Unlike Ethereum-based \ac{dexs} that work using smart contracts, the \ac{xrpl}-\ac{amm}-\ac{dex} is integrated directly at the protocol level of the \acl{xrpl}. It also has a special feature called \acl{cam} that seeks to reduce impermanent losses for \ac{lps} by giving them extra fees from traders who want to profit from price differences by participating in auctions to get a 24-hour zero-fee trading slot. This fundamental difference in infrastructure and features provides an interesting opportunity to analyze how these approaches affect \ac{amm}-based \ac{dexs} performance and characteristics.

Our study compares the \ac{xrpl}-\ac{amm}-\ac{dex} with a \ac{g-amm} based on Uniswap V2, which dominates 60\% of the \ac{dexs} market \cite{Lee2023}. While Uniswap V3 introduces concentrated liquidity in price ranges $[P_{a}, P_{b}]$, it behaves similarly to V2's $[0,\infty]$ distribution for trades within the same price range. Therefore, returns and losses scale with concentration, assuming the pool's current price stays within the same price range, and especially considering that most retail \ac{lps} often provide passive liquidity around current prices due to the challenges of active management in V3 \cite{Heimbach2022RisksProviders}. Given these similarities and theoretical considerations, our findings would likely apply to Uniswap V2 and V3 when benchmarking the \ac{g-amm} to the \ac{xrpl}-\ac{amm}-\ac{dex}.

Our methodology uses agent-based simulations\footnote{\label{note_xrpl_amm_simulation_code}https://github.com/dlt-science/xrpl-amm}, building on literature analyzing \ac{amm}-based \ac{dexs}' performance and design trade-offs \cite{Cohen2024InefficiencySimulations,Sabate-Vidales2023TheSimulation} and drawing from literature on the relationship between \ac{lps}' impermanent losses and traders' price slippage \cite{Evans2020LiquidityMarkets,Lehar2021DecentralizedExchanges,Capponi2021TheExchanges,Park2023TheMaking}, a relationship Milionis et al. \cite{Milionis2022AutomatedLoss-Versus-Rebalancing,Milionis2024ExtendedMakers} show applies to all \ac{amms}. Additionally, to the best of our knowledge, we conduct the first agent-based simulation of an auction mechanism for \ac{amm}-based \ac{dexs}, benchmarking the \ac{xrpl}-\ac{amm}-\ac{dex}'s \ac{cam} feature under various volatility scenarios. This builds on research into \ac{lps}' impermanent losses \cite{Fritsch2024MeasuringLiquidity,Loesch2021ImpermanentV3,Cohen2024InefficiencySimulations,Sabate-Vidales2023TheSimulation} and auction mechanism proposals and theoretical implications for \ac{amm}-based \ac{dexs} to reduce impermanent losses \cite{Adams2024Am-AMM:Maker,Canidio2023ArbitrageursResponse,josojo2022MEVResearch}. We simulate the underlying infrastructure of the \ac{g-amm} on Ethereum due to its \ac{defi} popularity, layer 1 blockchain status like the \acl{xrpl}, and role in popularizing smart contracts \cite{HernandezCruz2023EvolutionLiterature}, in which most \ac{amm}-based \ac{dexs} are built.

Our experimental results show that the \ac{xrpl}-\ac{amm}-\ac{dex}, leveraging the \acl{xrpl} infrastructure, reduces slippage, improves price synchronization with external markets, and enhances operational efficiency. These findings highlight the importance of shorter block confirmation times for \ac{amm}-based \ac{dexs}, aligning with Fritsch and Canidio's empirical findings \cite{Fritsch2024MeasuringLiquidity} and Milionis et al.'s theoretical modeling \cite{Milionis2023AutomatedFees}. Also, our experiments show that as volatility increases and arbitrage opportunities grow, the \ac{cam} feature in the \ac{xrpl}-\ac{amm}-\ac{dex} helps reduce \ac{lps}' impermanent losses by distributing additional fees from arbitrageurs' auctions. These results are consistent with theoretical proposals for other \ac{amm}-based \ac{dexs} auction mechanisms seeking to capture \ac{mev} value from arbitrageurs and redistribute it to \ac{lps} \cite{Adams2024Am-AMM:Maker,Canidio2023ArbitrageursResponse,josojo2022MEVResearch}.

\section{Related Work}
\label{sec:related_work}

\subsection{AMM-based DEX}
\ac{amm}-based \ac{dexs} have revolutionized \ac{defi}, providing innovative ways to exchange assets and provide liquidity \cite{xu2021dexAmm,Richardson2020e}. The \ac{cpmm} model, popularized by Uniswap V2 \cite{Adams2020}, forms the basis of many \ac{amm}-based \ac{dexs} \cite{xu2021dexAmm,Adams2020}. This model uses a simple bonding curve to set asset prices \cite{xu2021dexAmm}. As the field has grown, various \ac{amm}-based \ac{dexs} designs have emerged, each addressing specific market needs. These include Uniswap V3's concentrated liquidity, Balancer's multi-asset pools, Curve.fi's focus on similar-valued assets, and DODO's use of external price feeds \cite{xu2021dexAmm}. Despite these innovations, most \ac{amms} remain adaptations of the \ac{cpmm} model, highlighting its importance in \ac{defi} \cite{Adams2020,xu2021dexAmm}. 

A key challenge in \ac{amm}-based \ac{dex} design is balancing \MakeLowercase{\ac{lps}} and traders' interests. Milionis et al. \cite{Milionis2023AutomatedFees,Milionis2022AutomatedLoss-Versus-Rebalancing} show that \ac{lps}' impermanent losses stem from price slippage, as \ac{amms} only update prices during trades, unlike \ac{lobs} market makers who actively adjust quotes in response to buy and sell orders activity \cite{Amihud1986AssetSpread}. This limitation often results in suboptimal pricing, with trading fees frequently insufficient to offset \ac{lps} for arbitrage losses \cite{Fritsch2024MeasuringLiquidity,Loesch2021ImpermanentV3}, especially in Uniswap V3 \cite{Hasbrouck2023AnLiquidity}, where active liquidity management may disadvantage retail investors \cite{Heimbach2022RisksProviders}.

Proposed solutions include reducing block time to minimize arbitrage opportunities and associated \ac{lps} losses \cite{Fritsch2024MeasuringLiquidity}, implementing dynamic fee structures that adjust based on market volatility \cite{Sabate-Vidales2023TheSimulation}, and introducing governance mechanisms for fee adjustment, such as Uniswap V3's DAO voting system\footnote{https://gov.uniswap.org/t/uniswap-v3-fees-factory-owner-amendment/23187} and the \ac{xrpl}-\ac{amm}-\ac{dex}'s votable trading fee governance \cite{github}. Other innovations involve developing auction mechanisms for fee-setting rights \cite{Adams2024Am-AMM:Maker}, similar to traditional market-making practices in \ac{lobs} for setting bid-ask spreads according to volatility \cite{Kyle1985ContinuousTrading,Kyle1989InformedCompetition,Glosten1985BidTraders,GROSSMAN1988LiquidityStructure}, batch auctions \cite{Canidio2023ArbitrageursResponse}, and auctions to get the right for a trade to be placed first in a block of transactions \cite{josojo2022MEVResearch}. Among these auction mechanisms, the \ac{xrpl}-\ac{amm}-\ac{dex} proposes a \ac{cam} \cite{github} for auctioning daily zero-fee trading slots (\autoref{fig:slotPriceAlgo}).

\subsection{Performance evaluation}

\ac{amm}-based \ac{dexs} performance evaluation employs various techniques, with \ac{gbm} commonly generating price data for analyzing impermanent loss, price slippage, and price synchronization with an external market \cite{Milionis2022AutomatedLoss-Versus-Rebalancing,Cohen2024InefficiencySimulations,Sabate-Vidales2023TheSimulation}. Agent-based simulations examine \ac{lps}-trader dynamics \cite{Cohen2024InefficiencySimulations,Sabate-Vidales2023TheSimulation} with some research incorporating stochastic volatility \cite{Milionis2022AutomatedLoss-Versus-Rebalancing} or historical price data \cite{Fritsch2024MeasuringLiquidity}. Our methodology builds on this foundation by employing \ac{gbm} for stochastic price data generation and Binance data for realistic market conditions. However, unlike \cite{Milionis2022AutomatedLoss-Versus-Rebalancing}, we model fee-paying arbitrageurs. Additionally, we test the \ac{xrpl}-\ac{amm}-\ac{dex}'s \ac{cam} under specific volatility scenarios.

Most existing \ac{amm}-based \ac{dexs} research predominantly focuses on Ethereum-based \ac{dexs} \cite{xu2021dexAmm} because of Ethereum's pivotal role in the popularization of smart contracts \cite{HernandezCruz2023EvolutionLiterature}, in which most \ac{dexs} are built. These \ac{dexs} based on smart contracts execute on top of a blockchain via, most of the time, the \ac{evm}. This architecture often lags in transaction execution speed compared to native transactions \cite{xu2021dexAmm}, leading to increased slippage. Therefore, to the best of our knowledge, no other \ac{amm}-based \ac{dexs} exist at the protocol level of a blockchain, making this inaugural study particularly valuable for understanding the performance implications of protocol-level integration versus traditional smart contract implementations.

\section{\ac{amm}-based \ac{dex} on the \acf{xrpl}}
\label{sec:background}

\begin{figure}[bt]
    \centering
    \includegraphics[width=0.40\textwidth]{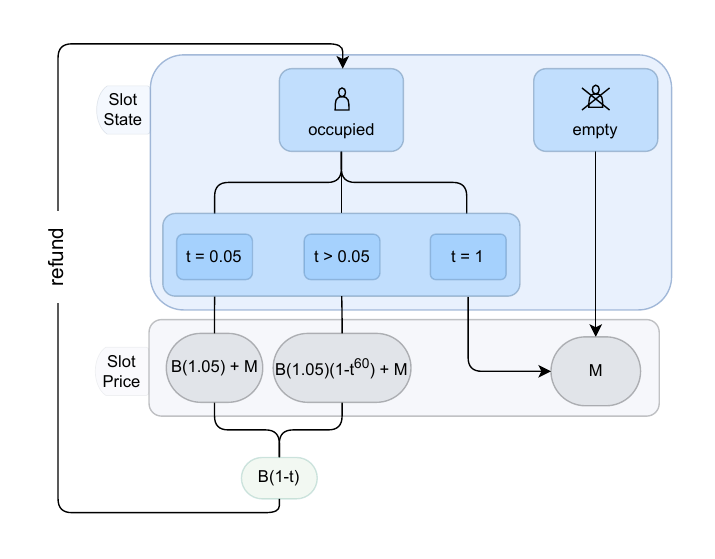}
    \captionsetup{font=scriptsize}
    \caption{\ac{xrpl}-\ac{amm}-\ac{dex}'s \ac{cam} slot price-schedule algorithm} \label{fig:slotPriceAlgo}
\end{figure}

We used the \ac{cpmm} at the core of our \acs{g-amm} for benchmark with the \ac{xrpl}-\ac{amm}-\ac{dex}'s \ac{g3m} constant product formula. The \ac{g3m} is similar to \ac{cpmm} for ensuring constant liquidity and enables algorithmic pricing based on the ratio of two tokens in the pool \cite{Evans2020LiquidityMarkets}:
\begin{equation}
    C = Q_A^{W_A} \times Q_B^{W_B}
\end{equation}

The reserves ($Q_A$ and $Q_B$) in the pool before and after each trade must have the same normalized weight ($W$) with $C$ remaining constant. The price at the current state of the pool (or time $t_0$) is the \ac{sp}, which is the slope of the conservation function or, to be more exact, the weighted ratio of the tokens $A$ and $B$ balances in the current state of the pool. The \ac{sp} also needs to incorporate the trading fee (up to $1\%$) of the liquidity pool ($\it TFee$), which is charged on the portion of the trade that changes the ratio of tokens in a pool. Then, the \ac{sp} is defined as:

\begin{equation}
    SP_{A}^{B} = \frac{\frac{Q_B}{W_B}}{\frac{Q_A}{W_A}} \times \frac{1}{1-{\it TFee}}
\end{equation}

Therefore, when a trade removes some amount of asset $A$ from the pool, they must put some amount of asset $B$ to preserve $C$ represented in the following swapping functions:

\paragraph{Swap Out} Minimum
    amount $\Delta_{B}$ of token $B$ to put into the pool to receive $\Delta_{A}$ amount of token $A$:
    
    {\begin{equation} \label{eq:swap_out}
    \Delta_{B} = Q_{B} \left[ ( \frac{Q_{A}}{Q_{A} - \Delta_{A}}) ^ \frac{W_A}{W_B}-1 \right] \frac{1}{1-{\it TFee}}           
    \end{equation}}

\paragraph{Swap In} Maximum amount of $\Delta_{A}$ to receive for paying $\Delta_{B}$ amount of token $B$:

    {\begin{equation} \label{eq:swap_in}
    \Delta_{A} = Q_{A} \left[1 - (\frac{Q_{B}}{Q_{B}+ \Delta_{B} \cdot (1-{\it TFee})})^\frac{W_{B}}{W_{A}}\right]   
    \end{equation}}

Particularly, the less liquidity there is in the pool, the more a single trade/swap may affect the price:

\begin{equation}
    \text{Price Impact} = \frac{\text{Price}_\text{post-swap}}{\text{Price}_\text{pre-swap}} - 1
\end{equation}

During the swapping of tokens $A$ and $B$, considering that the transaction is submitted at time $t_0$ but executed at time $t_1$, the actual executed price differs from the \acl{sp}. Therefore, the actual executed price or \ac{ep} of the trade is:

\begin{equation}
    EP_{A}^{B} = \frac{\Delta_{B}}{\Delta_{A}}
\end{equation}

This relationship between the \acl{sp} and \acl{ep} is the slippage, which may occur because of market movements between the delay ($t_0 - t_1$) when the trade transaction is submitted ($t_0$) versus finalized ($t_1$):

\begin{equation}
    \text{Slippage} = \frac{\text{\acl{ep}}}{\text{\acl{sp}}} - 1
\end{equation}
 
Slippage is one of the main \ac{mev} issues faced by \ac{amm}-based \ac{dexs} \cite{Eskandari2020SoK:Blockchain,Daian2020FlashInstability}. On the other hand, similar to traders experiencing slippage, \ac{lps} face impermanent loss from opportunity costs due to price volatility of their supplied assets in \ac{dexs}, with volatility significantly intensifying in \acp{dex} during market shocks \cite{HernandezCruz2024NoBank}.

\subsection{\acl{cam}} \label{subsec:cam}

The \ac{xrpl}-\ac{amm}-\ac{dex}'s \ac{cam} (\autoref{fig:slotPriceAlgo}) enables LPToken holders to bid for daily zero-fee trading slots, attracting arbitrageurs while maintaining standard access for other users. Winners retain slots until outbid or until the 24-hour period expires (\autoref{fig:slotPriceAlgo}).

\section{Testing Methodology}
\label{sec:methodology}

\begin{table}[tb]
\centering
\tiny %
\captionsetup{font=scriptsize}
\caption[Simulation parameters for different scenarios]{Simulation parameters for different scenarios: \ac{xrpl}-\ac{amm}-\ac{dex} vs. \ac{g-amm} (Test-1 and Test-2) and \ac{xrpl}-\ac{amm}-\ac{dex}'s \ac{cam}.}
\begin{tabular}{p{4cm} p{0.5cm} p{0.5cm} p{0.5cm}}
\toprule
\multicolumn{1}{c}{\textbf{Parameter}} & 
\multicolumn{1}{c}{\textbf{Test-1}} & 
\multicolumn{1}{c}{\textbf{Test-2}} & 
\multicolumn{1}{c}{\parbox[c]{1.8cm}{\centering\textbf{\ac{xrpl}-\ac{amm}-}\\\textbf{\ac{dex}'s \ac{cam}}}} \\
\midrule
\ac{xrpl} network fees (\ac{usdc}) & \textbf{1} & 0.00001 & 0.00001 \\
Ethereum network fees (\ac{usdc}) & \textbf{1} & 4 & 4 \\
\ac{xrpl} block interarrival time (seconds) & 4 & \textbf{8} & 4 \\
Ethereum block interarrival time (seconds) & 12 & \textbf{8} & 12 \\
Safe profit margin ($\%$) & 1.5 & 1.5 & 1.5 \\
Maximum slippage ($\%$) & 4 & 4 & 4 \\
\bottomrule
\end{tabular}
\label{tab:simulation_parameters}
\end{table}

\subsection{Data and environment modelling}
\label{subsec:environment_modelling}

 For our analysis, we choose the ETH/\ac{usdc} pair, a top-ten Uniswap pool based on \ac{tvl} \cite{coingeckoUniswap, UniswapPools}, using \ac{usdc} as numéraire. To evaluate the \ac{xrpl}-\ac{amm}-\ac{dex} (with and without its \ac{cam}) against a \ac{g-amm}, we combine two types of price data: 1) simulated data via \ac{gbm} and 2) real market data from Binance. Our goal is to test how each \ac{amm}-based \ac{dex} design responds to different market conditions for fee structures, block times, and market volatilities, focusing on price synchronization, \ac{lps} returns, and arbitrage metrics (slippage, profits).

\subsubsection{Simulated \ac{gbm} data}
We use \ac{gbm} to generate 5,000 price points over five days, starting with 1,000 \ac{usdc} per ETH, consistent with Black and Scholes \cite{Black1973TheLiabilities} 's groundbreaking options pricing model and Merton \cite{Merton1974ONRATES} 's application of it to corporate debt valuation modeling\cite{Black1973TheLiabilities,Merton1974ONRATES} while in \ac{amm}-based \ac{dexs} research, \ac{gbm} is used for analyzing impermanent loss, slippage, and price synchronization with an external market \cite{Milionis2022AutomatedLoss-Versus-Rebalancing,Cohen2024InefficiencySimulations,Sabate-Vidales2023TheSimulation,uniswapReturns}. \ac{gbm} is described by the formula: $S_t = S_0 \cdot e^{(\mu - \frac{\sigma^2}{2})t + \sigma  W_t}$, where $S_t$ is the price at time $t$, $S_0$ is the initial price, $\mu$ is the drift or expected return, $\sigma$ is the volatility of returns, $t$ is the time elapsed, and $W_t$ is a Wiener process, introducing random normal noise into the model. Following empirical evidence \cite{empirical_evidence,empirical_evidence2,empirical_evidence3}, we set the initial daily \ac{gbm} mean to 0.8\% and volatility to 7.7\%. Also, by adjusting drift ($\mu$) and volatility ($\sigma$), we can test the \ac{xrpl}-\ac{amm} with \ac{cam} under low, moderate, or high volatility. 

\subsubsection{Real market price data}

To confirm whether \ac{gbm}-like patterns hold in actual market conditions for the \ac{amm}-based \ac{dexs}, we also replicate our tests (\autoref{tab:simulation_parameters}) using five days of historical Binance ETH/\ac{usdc} prices\footnote{https://data.binance.vision/data/spot/daily/klines/ETHUSDC/1s/} (1-5 January 2024).

\subsubsection{Environment and tests}

We ran two tests on a shared reference market, excluding the \ac{cam} feature of the \ac{xrpl}-\ac{amm}-\ac{dex}. In \textbf{test-1}, both networks have a 1 \ac{usdc} fee, doubling the \ac{xrpl} minimum fee of 0.00001 \ac{usdc} used in \textbf{test-2} to anticipate fee fluctuations. In \textbf{test-2}, block times are equalized at eight seconds \cite{xrpl_blocktime,eth_blocktime}, while safe profit margin and maximum slippage values reflect realistic ranges (0.5\%--5\% \cite{uniswapSlippageRangeValues}). Next, we analyze the \ac{cam} feature of the \ac{xrpl}-\ac{amm}-\ac{dex} through two strategies (\autoref{arbitrageurs_strategies}): \ac{xrpl}-\ac{amm}-\ac{dex}-\ac{cam}-A (optimal for \ac{lps}) and \ac{xrpl}-\ac{amm}-\ac{dex}-\ac{cam}-B (optimal for arbitrageurs). We set $\mu=1\%$ for these simulations and choose three volatility levels (5\%, 12.5\%, and 20\%), simplifying each auction slot to a single user. \autoref{tab:simulation_parameters} summarizes the parameters.

\subsection{Agent-based simulation}

We adopt agent-based modeling\footnote{\label{note_xrpl_amm_simulation_code2}https://github.com/dlt-science/xrpl-amm} to examine how both \ac{amm}-based \ac{dexs} designs affect trading, liquidity, and price discovery under various conditions. This approach, common in Finance and Economics when modeling heterogeneous market participants who interact in stochastic and sometimes non-linear ways \cite{Bonabeau2002Agent-basedSystems,Farmer2009TheModelling}, captures behaviors like herd \cite{Bouri2019}, contrarian \cite{King2021}, and arbitrage strategies (particularly relevant for testing the \ac{xrpl}-\ac{amm}-\ac{dex} 's \ac{cam}). These interactions are often complex to capture in closed-form equilibrium models and can be obscured in live blockchain systems because of network congestion and delays, dynamic transaction fees, etc.

We simulate block interarrival times to approximate each \ac{dex}' underlying infrastructure, removing the frictions of smart contracts and other blockchain-specific limitations but focusing on core design differences, including the \ac{xrpl}-\ac{amm}-\ac{dex}'s \ac{cam} (\autoref{fig:slotPriceAlgo}). We set a $0.3\%$ trading fee, matching four of the top five Uniswap pools.\footnote{https://app.uniswap.org/explore/pools} Our two agents are:

\paragraph{Exchange users} Perform swap transactions, exchanging one asset for another. To reflect high market volatility and herd mentality \cite{Bouri2019}, users have an $80\%$ chance to trade ETH and a $20\%$ chance to abstain. Users are influenced by previous actions, with a $60\%$ probability of mimicking and $40\%$ of acting contrary, representing the mix of herd and contrarian behaviors in these markets \cite{King2021}. Order sizes range from $0.01$ to $2$ ETH.

\paragraph{Arbitrageurs}\label{arbitrageurs_strategies} Following rational arbitrage theory \cite{Shleifer1997TheArbitrage} with risk-adjusted profit targeting, these agents act as \blockquote{price balancers}, exploiting price differences between the \ac{amm} and external markets. They buy ETH or \ac{usdc} from the pool when prices diverge, aiming to sell for profit elsewhere. Their strategy involves: 1) Identify price difference: $|Price_{AMM} - Price_{ExternalMarket}| > 0$. 2) If a discrepancy exists, determine asset quantity for price alignment using equations \eqref{eq:swap_out} and \eqref{eq:swap_in}. 3) Compute potential profits by re-selling in the external market: ${\it Profits}_{\it potential} = \Delta_{\it assetOut}^- - \Delta_{\it assetIn}^+ - {\it networkFees}$. 4) Market microstructure research \cite{Stoll1978THEMARKETS,OHara2006MarketTheory} demonstrates that arbitrageurs need minimum profit margins to cover transaction costs and inventory risks, particularly in \ac{dexs}, where these costs are amplified by \ac{mev} competition \cite{Zhou2021High-FrequencyExchanges,Daian2020FlashInstability}. Therefore, arbitrageurs execute when risk-adjusted returns exceed a risk-premium threshold, named as \blockquote{Safe Profit Margin}:
$ \frac{\it Profits_{\it potential}}{\Delta_{assetIn}^+} > {\it SafeProfitMargin}
$

Then, the arbitrage condition can be expressed in Iverson bracket notation\footnote{The Iverson bracket notation denotes that $\llbracket P \rrbracket=1$ if the proposition $P$ is true and 0 otherwise.}:

\begin{equation}
\scriptsize
\begin{split}
\label{eq:arbitrage_condition_expanded}
\llbracket |Price_{AMM} - & Price_{ExternalMarket}| > 0 \rrbracket \cdot \\
& \llbracket \frac{\Delta_{assetOut}^- - \text{networkFees}}{\Delta_{assetIn}^+} - 1 > \text{SafeProfitMargin} \rrbracket = 1
\end{split}
\nonumber
\end{equation}

In addition to the above, arbitrageurs on the \ac{xrpl}-\ac{amm}-\ac{dex}-\ac{cam} (\ac{xrpl}-\ac{amm}-\ac{dex} with \ac{cam}) can bid for the discounted trading fee based on two distinct strategies:

\subsubsection{Case A: \ac{xrpl}-\ac{amm}-\ac{dex}-\ac{cam}-A} \label{subsubsec:case-a}

This scenario favors liquidity providers over arbitrageurs, but their interaction is more complex than a zero-sum game. Arbitrageurs often bid for and hold slots for entire blocks. The simulation starts on day three, providing arbitrageurs with historical data to estimate profits under a $0\%$ trading fee scenario. This approach aligns historical and simulated data at $S_0$ ($1000$ \ac{usdc}/ETH), mimicking real market conditions where traders may use the available information (including past) to guide their strategies. So, the weighted average bid limit, $P$, is determined using exponential smoothing to prioritize recent data. Arbitrageurs cap their bids at $P$ and adjust them based on daily profit trends until the minimum bid price, $M$, exceeds $P$. They calculate the LPToken value relative to \ac{usdc} as follows:

\begin{equation} \label{eq:lpt_price}
\tiny
LPToken_{RelativePrice}  = \frac{SP_{A}^{B} \cdot Q_A + Q_B}{Q_{LPTokens}}
\nonumber
\end{equation}
where $A=ETH$ and $B=\ac{usdc}$. The expected outcomes from this strategy are (a) decreased arbitrageurs profits and (b) increased \ac{lps} returns.

\subsubsection{Case B: \ac{xrpl}-\ac{amm}-\ac{dex}-\ac{cam}-B} \label{subsubsec:case-b}

This scenario favors arbitrageurs over liquidity providers, with minimal competition. An arbitrageur secures the slot at the minimum bid $M$ when empty (\autoref{fig:slotPriceAlgo})
and controls it for 24 hours, repeating until the simulation ends. The anticipated outcomes from this structure are (a) maximal profits for arbitrageurs and (b) minimal returns for liquidity providers.

\subsubsection{Number of arbitrageurs}

In Case A (\autoref{subsubsec:case-a}), arbitrageurs bid continuously until $M>P$, yielding the same outcome regardless of arbitrageur count. In Case B (\autoref{subsubsec:case-b}), daily bidding produces equivalent results, whether from multiple arbitrageurs or one renewing daily. Still, we conducted simulations with varying numbers, consistently obtaining similar results. We settled on using five arbitrageurs in our final simulations.

\subsection{Set up}

All scenarios begin with initial pool reserves of 50,000 ETH and 49,850,000 \ac{usdc}, setting the initial ETH price for \ac{gbm} pricing at 1,000 \ac{usdc} to match the external market price ($S_0$) and 2,281.57 \ac{usdc} using real market price data. \autoref{tab:simulation_parameters} summarizes key parameters: network fees, block times, safe profit margin, and so on, used to compare the \ac{xrpl}-\ac{amm}-\ac{dex} (with and without \ac{cam}) and the \ac{g-amm}. We repeat the simulations on both \ac{gbm} (multiple volatility levels) and Binance data.

\section{Results}
\label{sec:results}

Given that the results for arbitrageur profits, \ac{lps} returns, and impermanent loss are nearly identical, we consolidated \ac{xrpl}-\ac{amm}-\ac{dex}-\ac{cam}-A and B as \ac{xrpl}-\ac{cam} for clarity in figures and reports. Similarly, we abbreviate \ac{xrpl}-\ac{amm}-\ac{dex} to \ac{xrpl}-\ac{amm} using both terms interchangeably, but both referring to the \ac{amm}-based \ac{dex} in the \acl{xrpl}. Results are averages from multiple simulations due to random transaction processing, the results of which vary slightly between tests. While specific values may vary, the key insights lie in the relative performance differences between the \ac{amm}-based \ac{dexs} across various market scenarios with simulated and historical price data.

\begin{figure*}[t]
    \centering
    \begin{minipage}[t]{0.20\textwidth}
        \centering
        \includegraphics[width=\linewidth]{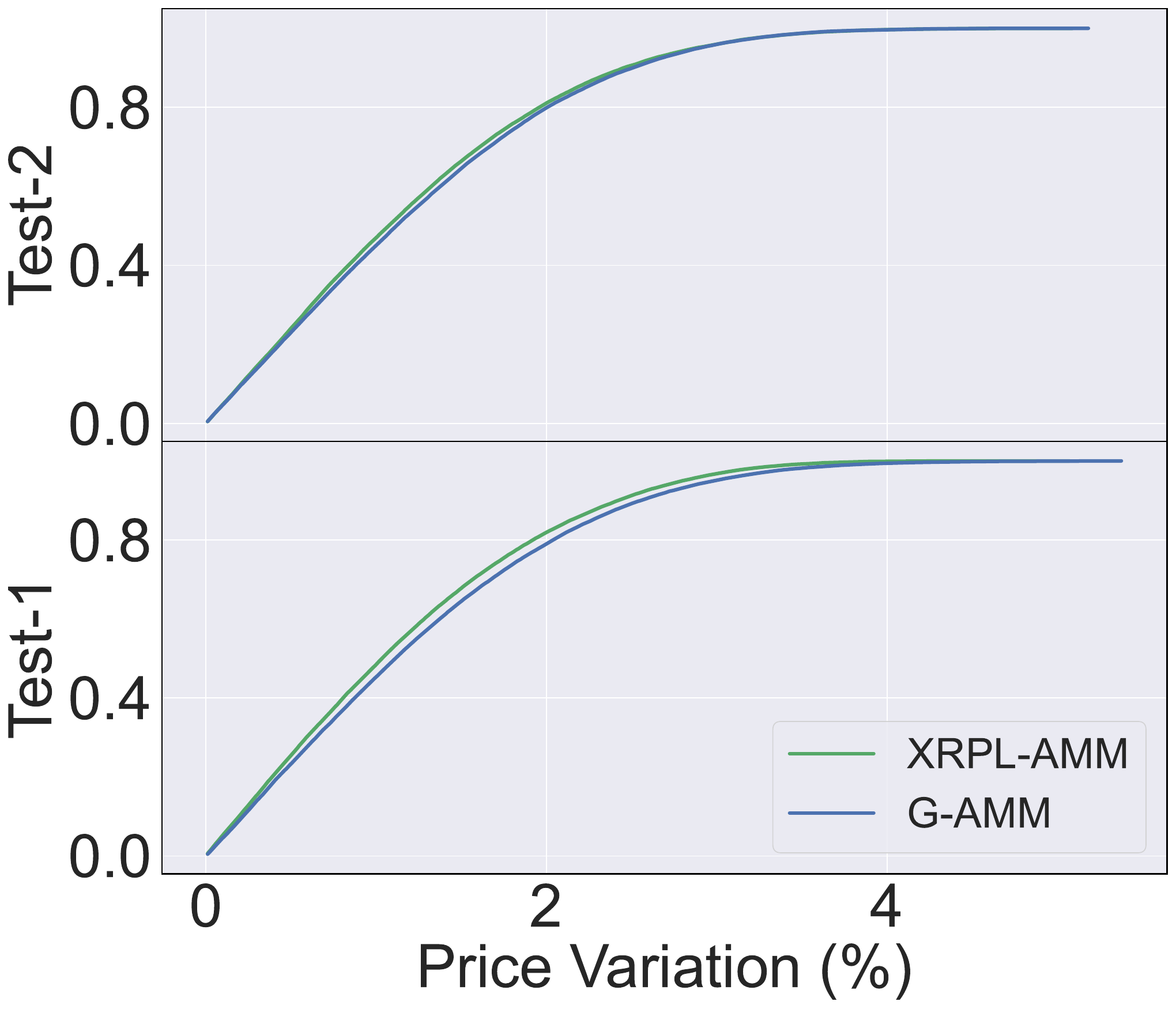}
        \captionsetup{font=scriptsize}
        \caption[Price Sync. with Reference Market for \ac{xrpl}-\ac{amm}-\ac{dex} vs. \ac{g-amm} (\ac{cdf})]{Test-1 and Test-2 Cumulative Distribution Functions (CDFs) of the price sync. with the reference market for \ac{xrpl}-\ac{amm}-\ac{dex} vs. \ac{g-amm}.}
        \label{fig:xrpl_vs_uniswap_price_sync}
    \end{minipage}
    \hfill
    \begin{minipage}[t]{0.20\textwidth}
        \centering
        \includegraphics[width=\linewidth]{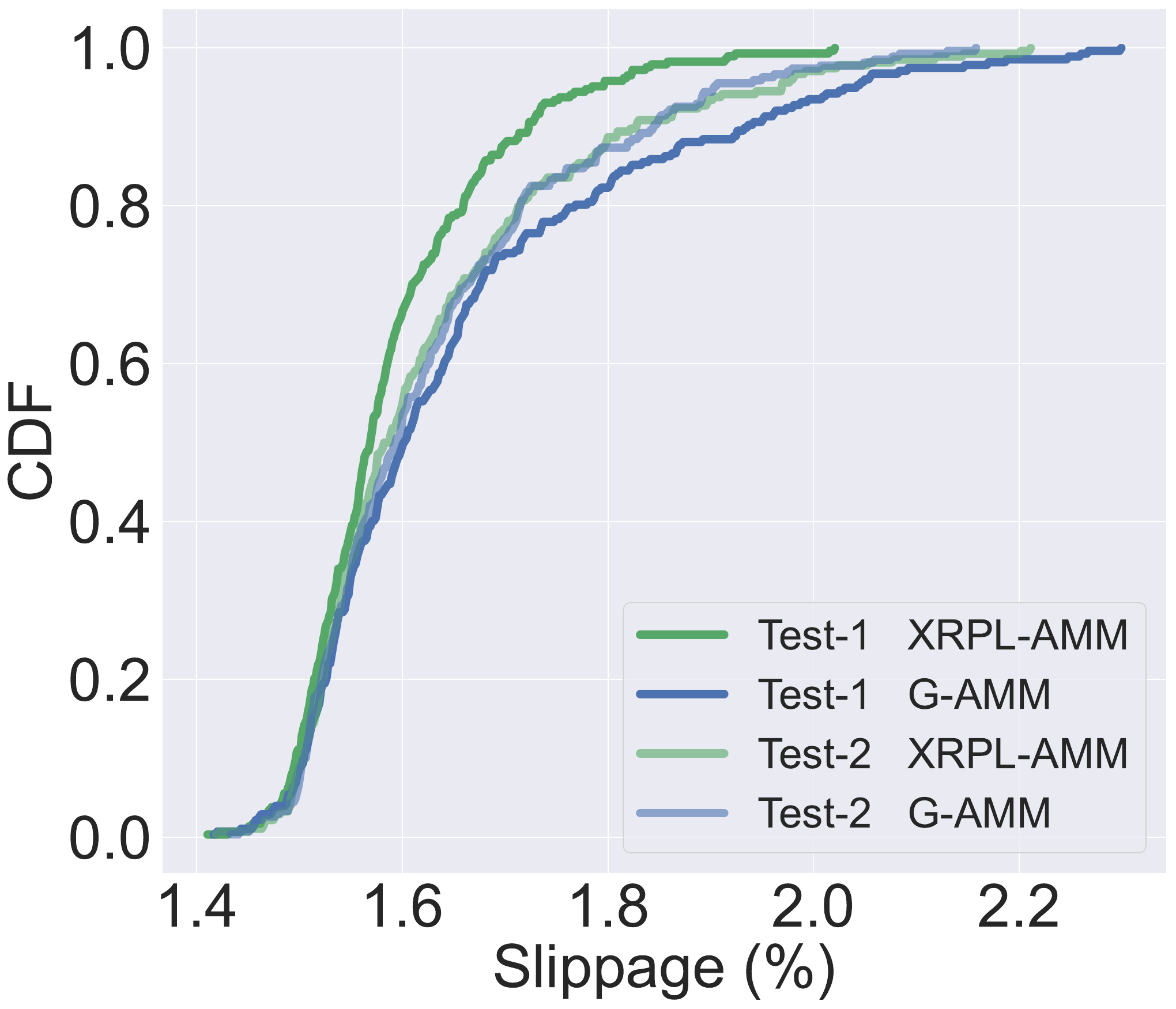}
        \captionsetup{font=scriptsize}
        \caption[Slippage for \ac{xrpl}-\ac{amm}-\ac{dex} vs \ac{g-amm} (\ac{cdf})]{Test-1 and Test-2 \ac{cdfs} of the slippage.}
        \label{fig:xrpl_vs_uniswap_slippages}
    \end{minipage}
    \hfill
    \begin{minipage}[t]{0.20\textwidth}
        \centering
        \includegraphics[width=\linewidth]{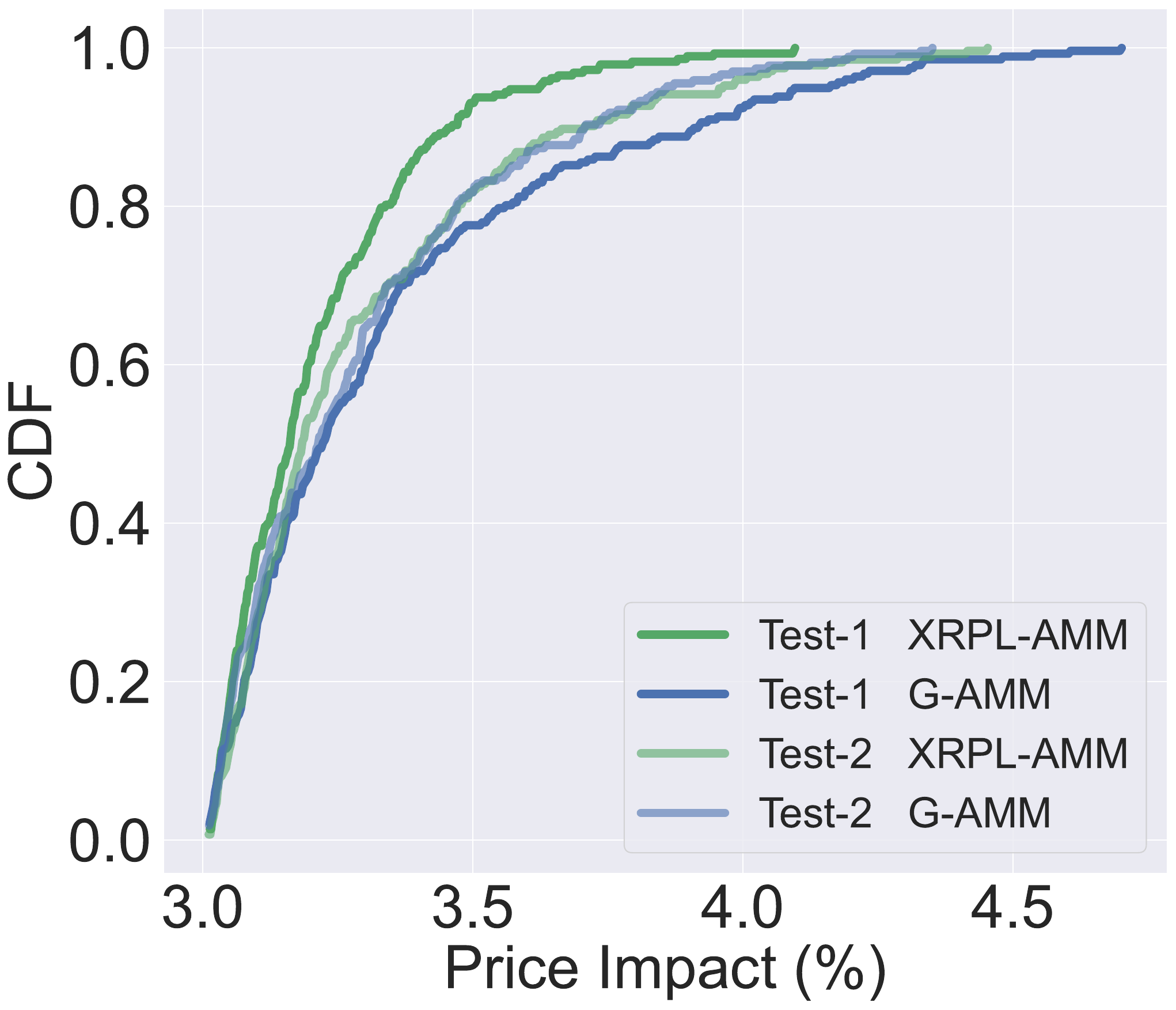}
        \captionsetup{font=scriptsize}
        \caption[Price Impact for \ac{xrpl}-\ac{amm}-\ac{dex} vs. \ac{g-amm} (\ac{cdf})]{Test-1 and Test-2 CDFs of the price impact.}
        \label{fig:xrpl_vs_uniswap_price_impact_cdf}
    \end{minipage}
    \hfill    
    \begin{minipage}[t]{0.31\textwidth}
        \centering
        \includegraphics[width=\linewidth]{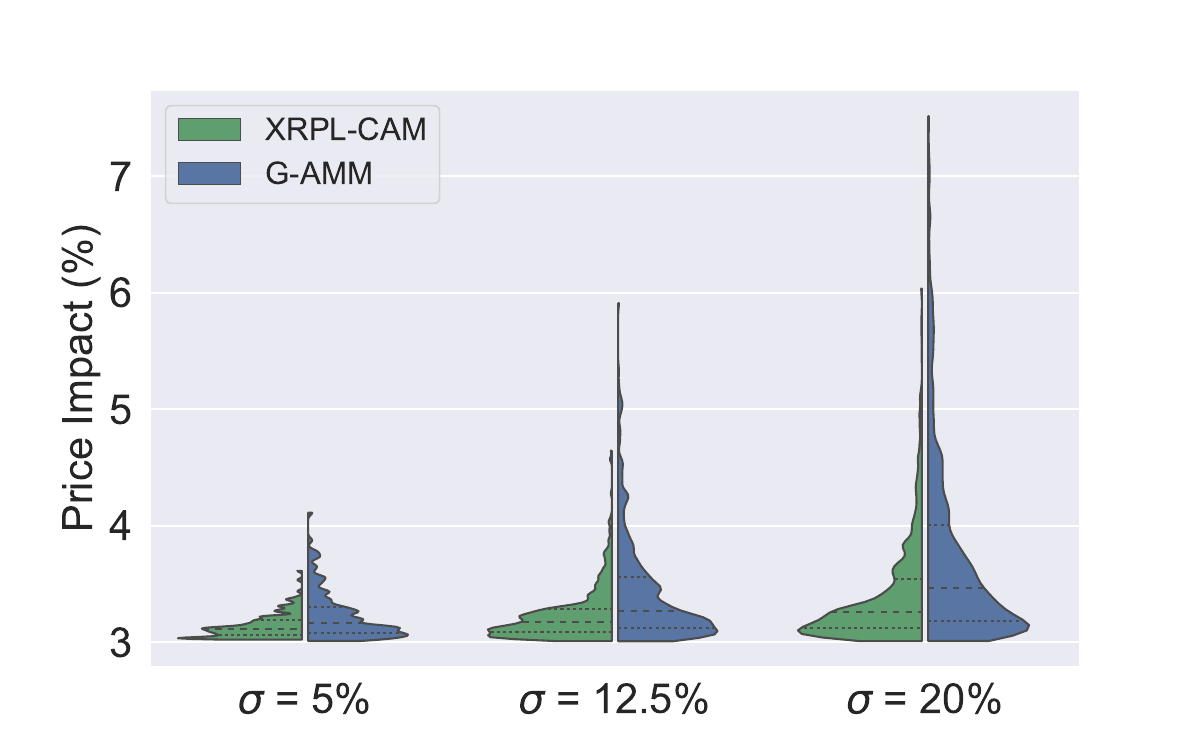}
        \captionsetup{font=scriptsize}
        \caption[Price Impact for \ac{xrpl}-\ac{cam} vs \ac{g-amm} with different volatilities (Violin Plot)]{Price impact caused by arbitrageurs under three varying volatilities. The dotted lines represent the first, second (median), and third quartiles.}
        \label{fig:xrplCAM_vs_uniswap_price_impact_violin_plot}
    \end{minipage}
\end{figure*}

\subsubsection{Trading Volume}
Trading volumes\footnote{\label{note_normal_users_values_included}All analyses throughout the paper include normal users' trading volume and fees. Their difference is negligible as identical transactions are simultaneously placed on both \ac{amm}-based \ac{dexs}.} increase with market volatility for both \ac{xrpl}-\ac{cam} and \ac{g-amm}, with \ac{xrpl}-\ac{cam} consistently outperforming \ac{g-amm} by an average of 4\% across all volatility regimes. At $\sigma=5\%$, \ac{xrpl}-\ac{cam} registered 157,979,186 \ac{usdc} versus \ac{g-amm}'s 155,723,926 \ac{usdc}; at $\sigma=12.5\%$, volumes increased to 191,209,656 versus 182,497,061 \ac{usdc}; and at $\sigma=20\%$, trading activity escalated substantially to 289,189,151 versus 273,181,521 \ac{usdc}, respectively.

In test-1 and test-2, trading volumes show remarkable similarity across both \ac{amm}-based \ac{dexs} using simulated and historical price data. With equalized network fees and different block interarrival times (test-1), \ac{xrpl}-\ac{amm}-\ac{dex} volume was 170,746,887 \ac{usdc} versus \ac{g-amm}'s 170,721,881 \ac{usdc}, a $0.015\%$ difference. With varied fees and equalized block interarrival times (test-2), the difference increased to 0.23\%: \ac{xrpl}-\ac{amm}-\ac{dex} at 170,277,799 \ac{usdc} and \ac{g-amm} at 169,890,854 \ac{usdc}.

\subsubsection{Price Synchronization}
\ac{xrpl}-\ac{cam} consistently outperforms \ac{g-amm} in price alignment across all volatility levels. Comparing 80th percentile price gaps at $\sigma=5\%$, \ac{xrpl}-\ac{cam} achieves $1.7\%$ versus \ac{g-amm}'s $1.9\%$ ($11.8\%$ difference); at $\sigma=12.5\%$, the gap widens to $1.9\%$ versus $2.3\%$ ($21\%$ difference); and at $\sigma=20\%$, this divergence further amplifies to $2.1\%$ versus $2.7\%$ ($28.6\%$ difference).

\begin{figure*}[tb]
    \centering
    \begin{minipage}[t]{0.28\textwidth}
        \centering
        \includegraphics[width=0.9\linewidth]{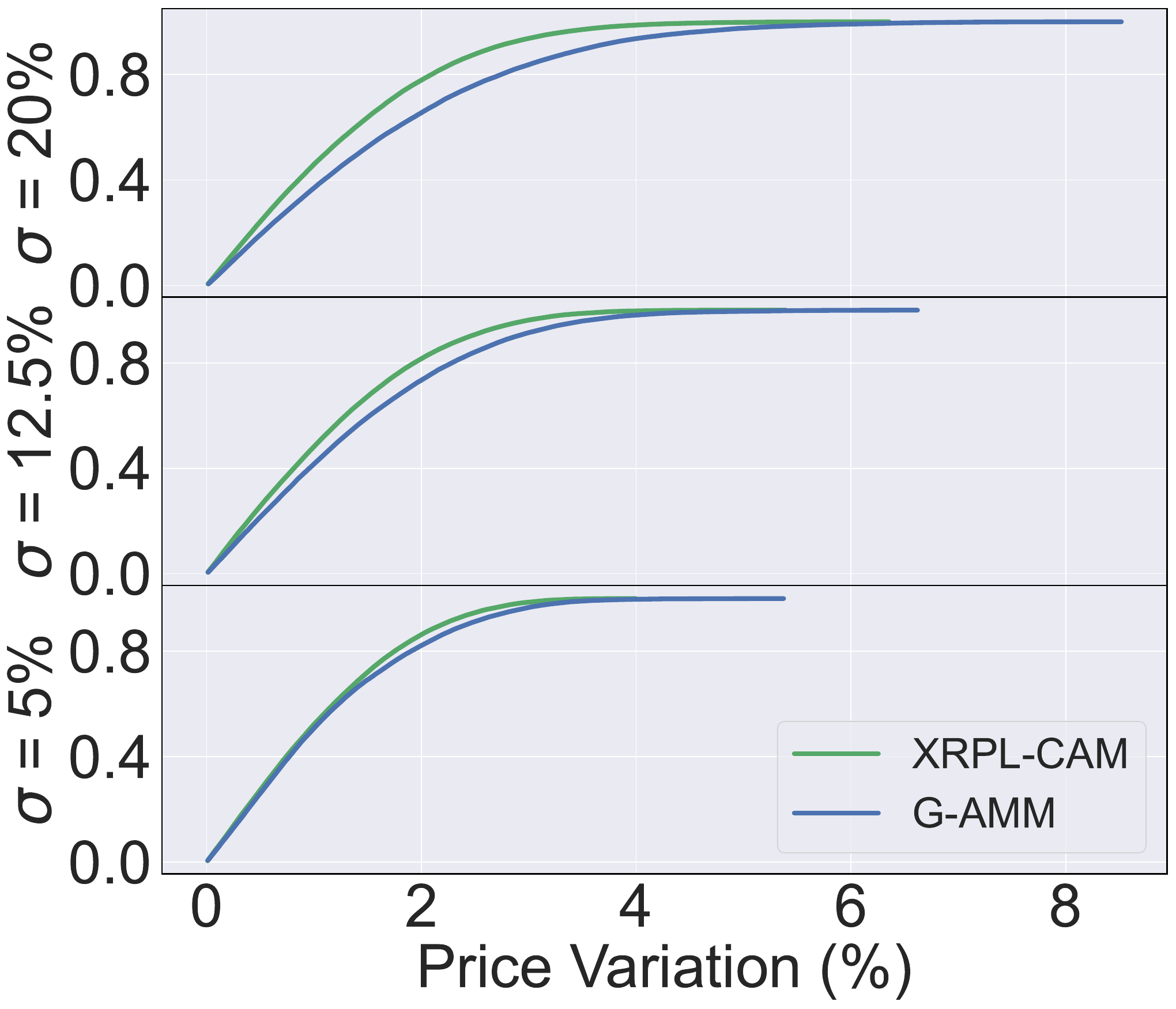}
        \captionsetup{font=scriptsize}
        \caption[Price Sync. for \ac{xrpl}-\ac{cam} vs \ac{g-amm} with different volatilities (\ac{cdf})]{\ac{cdfs} of the price difference with the external market for three volatilities.}
        \label{fig:xrplCAM_vs_uniswap_price_sync_cdf}
    \end{minipage}
    \hfill
    \begin{minipage}[t]{0.28\textwidth}
        \centering
        \includegraphics[width=1\linewidth]{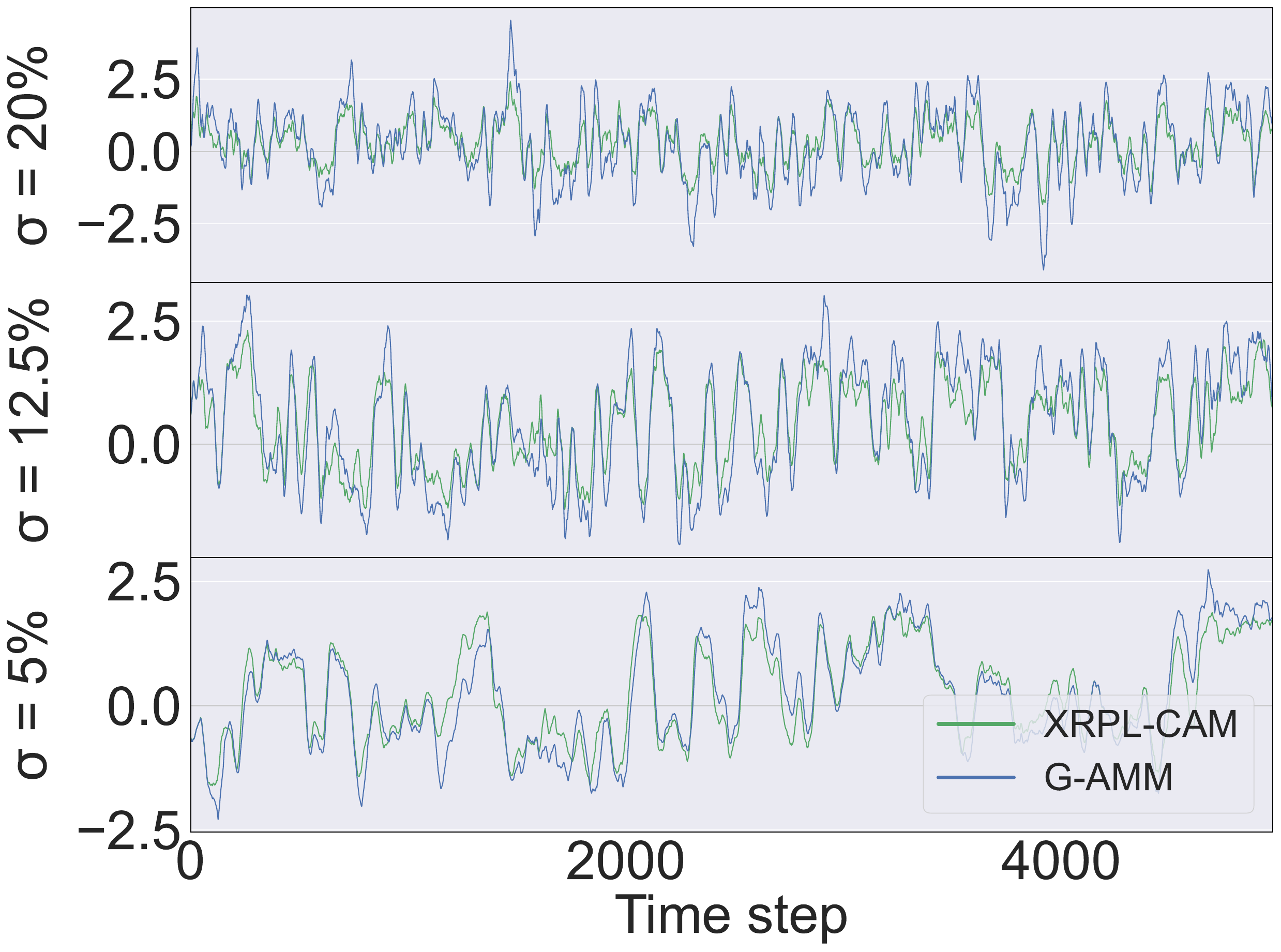}
        \captionsetup{font=scriptsize}
        \caption[Price Sync. for \ac{xrpl}-\ac{cam} vs. \ac{g-amm} with different volatilities (Moving Average)]{30-period moving averages of price discrepancies (\%) with external markets across three volatilities.}
        \label{fig:xrplCAM_vs_uniswap_price_sync_MA}
    \end{minipage}
    \hfill
    \begin{minipage}[t]{0.28\textwidth}
        \centering
        \includegraphics[width=0.9\linewidth]{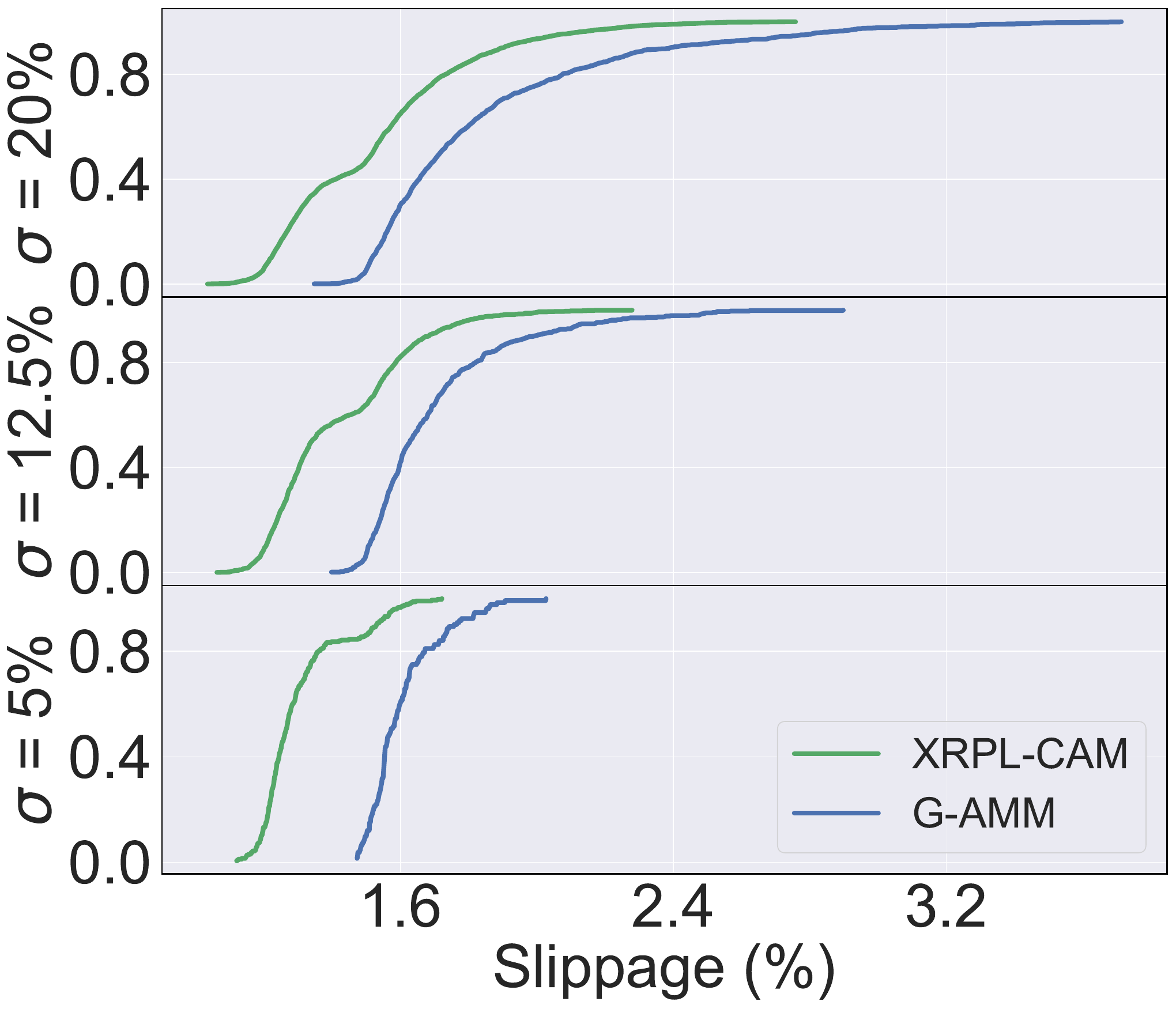}
        \captionsetup{font=scriptsize}
        \caption[Slippage for \ac{xrpl}-\ac{cam} vs. \ac{g-amm} with different volatilities (CDF)]{CDFs of the slippage for three volatilities.}
        \label{fig:xrplCAM_vs_uniswap_slippage}
    \end{minipage}
\end{figure*}

Using a moving average, \ac{xrpl}-\ac{cam} shows superior stability, never exceeding $2.4\%$ divergence across all scenarios, versus \ac{g-amm}'s $4.5\%$. \autoref{fig:xrplCAM_vs_uniswap_price_sync_MA} illustrates this trend. In test-1, with equal network fees but different block interarrival times, \ac{xrpl}-\ac{amm}-\ac{dex} outperformed \ac{g-amm} in $90\%$ of cases. In test-2, this advantage dropped to $60\%$ with equal block interarrival times but different fees (\autoref{fig:xrpl_vs_uniswap_price_sync}).

\begin{table}[tb]
\centering
\tiny %
\captionsetup{font=scriptsize}
\caption[Average Arbitrageurs' Profits, Transaction Costs \& Transaction Count for \ac{xrpl}-\ac{cam} vs. \ac{g-amm} with different volatilities]{Average arbitrageurs' profits, transaction costs \& transaction frequency for \ac{xrpl}-\ac{cam} vs. \ac{g-amm} with different volatilities.}

\begin{tabular}{cccccc}
\toprule
\multicolumn{2}{c}{\multirow{2}{*}{\textbf{Volatility}}} & \multirow{2}{*}{\textbf{Profits (\ac{usdc})}} & \multirow{2}{*}{\textbf{Fees (\ac{usdc})}} & \multicolumn{2}{c}{\textbf{Transaction Count}} \\
\cmidrule{5-6}
\multicolumn{2}{c}{} & & & \textbf{Realized (\%)} & \textbf{Unrealized} \\
\midrule
\multirow{3}{*}{$\sigma=5\%$}
& \ac{xrpl}-\ac{cam}-A & 97,251 & \multirow{2}{*}{0.0002} & \multirow{2}{*}{16 (31.4\%)} & \multirow{2}{*}{35} \\
& \ac{xrpl}-\ac{cam}-B & 180,303 & & & \\
& \ac{g-amm} & 174,686 & 53 & 13 (4\%) & 311 \\
\midrule
\multirow{3}{*}{$\sigma=12.5\%$}
& \ac{xrpl}-\ac{cam}-A & 235,937 & \multirow{2}{*}{0.001} & \multirow{2}{*}{72 (18.3\%)} & \multirow{2}{*}{322} \\
& \ac{xrpl}-\ac{cam}-B & 823,910 & & & \\
& \ac{g-amm} & 760,056 & 230 & 58 (4.8\%) & 1,150 \\
\midrule
\multirow{3}{*}{$\sigma=20\%$}
& \ac{xrpl}-\ac{cam}-A & 468,500 & \multirow{2}{*}{0.002} & \multirow{2}{*}{159 (15.4\%)} & \multirow{2}{*}{875} \\
& \ac{xrpl}-\ac{cam}-B & 2,159,411 & & & \\
& \ac{g-amm} & 1,985,052 & 512 & 128 (4.2\%) & 2,938 \\
\bottomrule
\end{tabular}
\label{tab:xrplCAM_uniswap_profits_fees_txCount}
\end{table}

\subsubsection{Price Impact}

Price impact increases with market volatility for all \ac{amm}-based \ac{dexs}, with the gap between \ac{xrpl}-\ac{cam} and \ac{g-amm} widening at higher volatilities (\autoref{fig:xrplCAM_vs_uniswap_price_impact_violin_plot} illustrates these trends). At $\sigma=5\%$, both mechanisms show similar average impacts, though \ac{g-amm} exhibits more outliers. At $\sigma=12.5\%$, \ac{xrpl}-\ac{cam} maintains a lower, more consistent mean price impact. This divergence amplifies at $\sigma=20\%$, where \ac{g-amm}'s mean price impact exceeds \ac{xrpl}-\ac{cam} by 10.3\%, indicating superior price stability in the latter mechanism under elevated market volatility.

In test-1, \ac{xrpl}-\ac{amm}-\ac{dex} showed less price impact than \ac{g-amm} with equal network fees and varying block time: $80\%$ of values remained below $3.3\%$ for \ac{xrpl}-\ac{amm}-\ac{dex}, versus $3.55\%$ for \ac{g-amm} -- a $7.6\%$ difference. Test-2, with equal block times and different fees, shows similar price impact distributions for both \ac{amm}-based \ac{dexs}. \autoref{fig:xrpl_vs_uniswap_price_impact_cdf} confirms this, with overlapping \ac{cdf} curves in test-2.

\subsubsection{Slippage}

\ac{xrpl}-\ac{cam} consistently shows lower slippage than \ac{g-amm}. Comparing 80th percentile slippage values at $\sigma=5\%$, \ac{xrpl}-\ac{cam} records $1.36\% $versus \ac{g-amm}'s $1.67\%$ ($22.3\%$ difference); at $\sigma=12.5\%$, values of $1.58\%$ versus $1.82\%$ ($15.2\%$ difference) are observed; and at $\sigma=20\%$, the disparity reaches $1.73\%$ versus $2.07\%$ ($19.7\%$ difference).

For test-1 and test-2, \ac{cdfs} (\autoref{fig:xrpl_vs_uniswap_slippages}) show that with equal interarrival block times, slippage is nearly identical on both \ac{amms}. However, with realistic block times for the \acl{xrpl} and Ethereum, \ac{xrpl}-\ac{amm}-\ac{dex} exhibits less slippage. In test-1, $80\%$ of slippages on \ac{g-amm} approach just below $1.8\%$, while on \ac{xrpl}-\ac{amm}-\ac{dex}, they are around $1.65\%$ -- an $8.8\%$ reduction.

\subsubsection{Impermanent/divergent Loss}

\begin{figure}[tb]
\begin{center}
\includegraphics[width=0.42\textwidth]{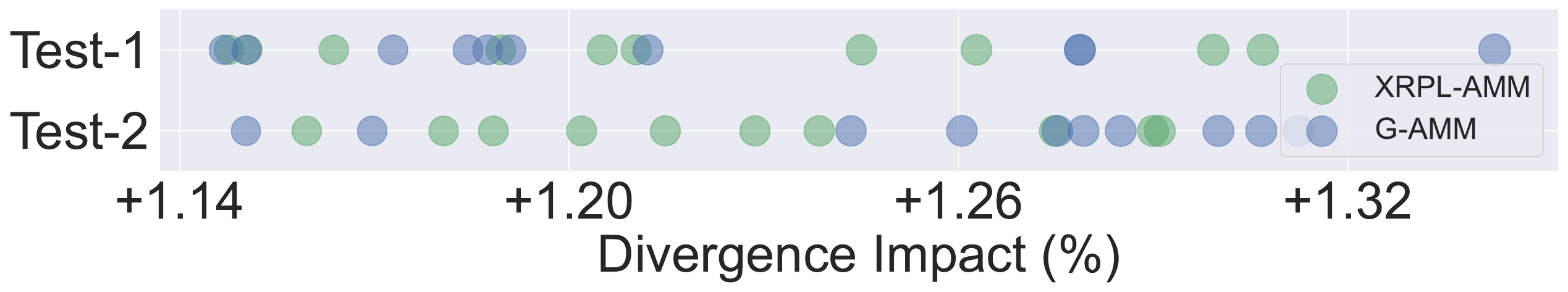}
\captionsetup{font=scriptsize}
\caption[Divergence Loss for \ac{xrpl}-\ac{amm}-\ac{dex} vs. \ac{g-amm} (Scatter Plot)]{\ac{lps}' divergence gains for Test-1 and Test-2 accross \ac{xrpl}-\ac{amm}-\ac{dex} and \ac{g-amm}.}
\label{fig:xrpl_vs_uniswap_divergence_loss}
\end{center}
\end{figure}

\begin{figure}[hbt]
\begin{center}
    \includegraphics[width=0.45
    \textwidth]{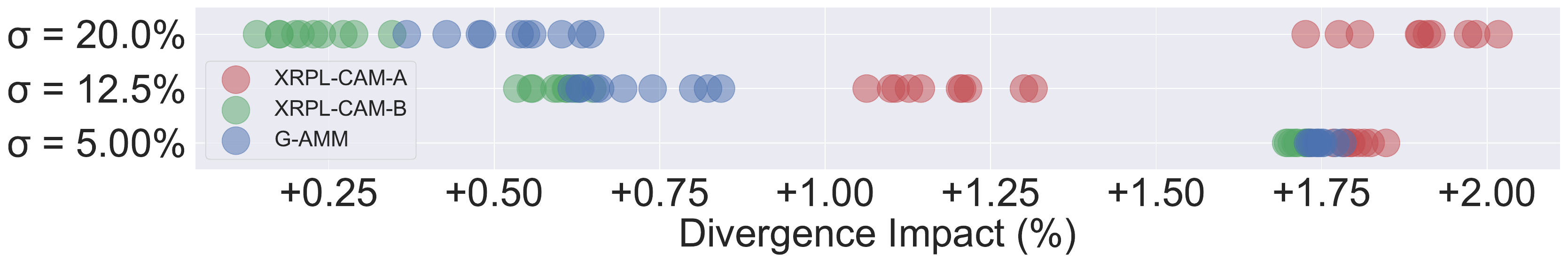}
    \captionsetup{font=scriptsize}
    \caption[Divergence Loss for \ac{xrpl}-\ac{cam} vs. \ac{g-amm} with different volatilies (Scatter Plot)]{Divergence loss/gain for \ac{xrpl}-\ac{cam} vs. \ac{g-amm} with different volatilities.} \label{fig:divergence_loss_xrplCAM_uniswap}
\end{center}
\end{figure}

\begin{figure}[tb]
\begin{center}
\includegraphics[width=0.40\textwidth]{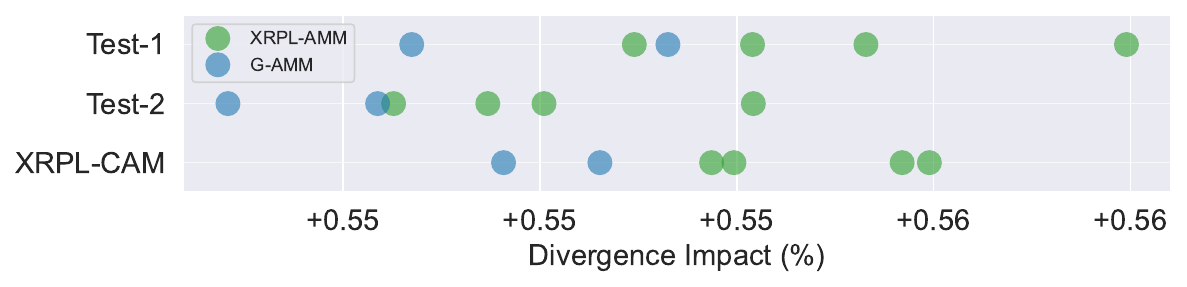}
\captionsetup{font=scriptsize}
\caption{\ac{lps}' divergence gains for \ac{xrpl}-\ac{amm}-\ac{dex} (with and without \ac{cam}) vs. \ac{g-amm} (\ac{cdf}) using historical price data from Binance}
\label{fig:xrpl_vs_uniswap_divergance_binance}
\end{center}
\end{figure}

\begin{table}[tb]
\centering
\tiny
\setlength{\tabcolsep}{4pt}  %
\captionsetup{font=scriptsize}
\caption[Avg. LPs' Returns: \ac{xrpl}-\ac{cam} vs. \ac{g-amm} under diff. volatilities]{Average LPs' returns under different volatilities for \ac{xrpl}-\ac{cam} vs. \ac{g-amm}.}
\begin{tabular}{@{}lcrrr@{}}  %
\toprule
\multirow{2}{*}{\textbf{Volatility}} & \multirow{2}{*}{\textbf{Scenario}} & \multicolumn{3}{c}{\textbf{Returns (\ac{usdc})}} \\
\cmidrule(lr){3-5}
& & \textbf{\ac{cam} Bids} & \textbf{Trading Fees} & \textbf{Total} \\
\midrule
\multirow{3}{*}{$\sigma=5\%$}
& \ac{xrpl}-\ac{cam}-A & 96,528 (10\%) & 875,883 (90\%) & 972,411 \\
& \ac{xrpl}-\ac{cam}-B & 12,233 & 877,947 & 890,180 \\
& \ac{g-amm} & -- & 900,401 & 900,401 \\
\midrule
\multirow{3}{*}{$\sigma=12.5\%$}
& \ac{xrpl}-\ac{cam}-A & 526,500 (38\%) & 870,565 (62\%) & 1,397,065 \\
& \ac{xrpl}-\ac{cam}-B & 10,269 & 871,100 & 881,369 \\
& \ac{g-amm} & -- & 948,282 & 948,282 \\
\midrule
\multirow{3}{*}{$\sigma=20\%$}
& \ac{xrpl}-\ac{cam}-A & 1,980,951 (64\%) & 1,113,244 (36\%) & 3,094,195 \\
& \ac{xrpl}-\ac{cam}-B & 14,566 & 1,104,358 & 1,118,924 \\
& \ac{g-amm} & -- & 1,264,858 & 1,264,858 \\
\bottomrule
\end{tabular}
\label{tab:returns_xrplCAM_uniswap}
\end{table}

\autoref{tab:returns_xrplCAM_uniswap} shows \ac{lps}' returns increase with market volatility, primarily from \ac{cam} bids and trading fees. Despite lower trading fee returns, \ac{xrpl}-\ac{cam}-A outperforms \ac{xrpl}-\ac{cam}-B and \ac{g-amm} in high volatility. \ac{xrpl}-\ac{cam}-A achieves $67\%$ higher returns than \ac{g-amm}, while \ac{xrpl}-\ac{cam}-B exhibits marginally inferior performance with returns $7.25\%$ below those of \ac{g-amm}. This differential performance underscores the protocol-specific sensitivity to volatility regimes and liquidity dynamics.

\ac{cam} contributions to total returns in \ac{xrpl}-\ac{cam}-A to reach $64\%$ at $\sigma=20\%$, indicating aggressive arbitrageur bidding in volatile markets, taking advantage of price fluctuations. At all volatility levels, \ac{lps} in \ac{xrpl}-\ac{cam}-A consistently show more divergence gain than \ac{xrpl}-\ac{cam}-B while \ac{g-amm} experienced divergence loss:
\begin{itemize}
\item $\sigma=5\%$: \ac{xrpl}-\ac{cam}-A $+1.8\%$, \ac{g-amm} $+1.75\%$, \ac{xrpl}-\ac{cam}-B $+1.72\%$
\item $\sigma=12.5\%$: \ac{xrpl}-\ac{cam}-A $+1.18\%$, \ac{g-amm} $+0.7\%$, \ac{xrpl}-\ac{cam}-B $+0.6\%$
\item $\sigma=20\%$: \ac{xrpl}-\ac{cam}-A $+1.9\%$, \ac{g-amm} $-0.5\%$, \ac{xrpl}-\ac{cam}-B $+0.2\%$
\end{itemize}

\autoref{fig:divergence_loss_xrplCAM_uniswap} illustrates growing disparities between \ac{xrpl}-\ac{cam}-A and others as volatility increases. \ac{xrpl}-\ac{cam}-B closely mirrors \ac{g-amm}, indicating comparable worst-case scenarios for \ac{lps}. 

For test-1 and test-2, \ac{lps} on \ac{xrpl}-\ac{amm}-\ac{dex} outperformed \ac{g-amm} in 7 out of 10 simulations, with a marginal $0.35\%$ advantage. In test-1, \ac{xrpl}-\ac{amm}-\ac{dex}'s \ac{lps} earned average returns of 954,394 \ac{usdc} with a $+1.22\%$ divergence gain, while \ac{g-amm} yielded 951,561 \ac{usdc} returns and a $+1.21\%$ divergence gain. Test-2 showed \ac{xrpl}-\ac{amm}-\ac{dex}'s \ac{lps} achieving 951,984 \ac{usdc} returns with a $+1.23\%$ divergence gain, compared to \ac{g-amm}'s 948,364 \ac{usdc} returns and $+1.26\%$ divergence gain. Despite similar overall results, test-2 revealed a $2.4\%$ higher divergence gain for \ac{lps} on \ac{g-amm}. \autoref{fig:xrpl_vs_uniswap_divergence_loss} depicts divergence gain distributions for both tests. Test-1 shows minimal difference, while test-2 reveals a skew to the right for \ac{g-amm} values compared to \ac{xrpl}-\ac{amm}-\ac{dex}, indicating marginally higher divergence gains.

\begin{figure*}[t]
    \centering
    \begin{minipage}[t]{0.28\textwidth}
        \centering
        \includegraphics[width=\linewidth]{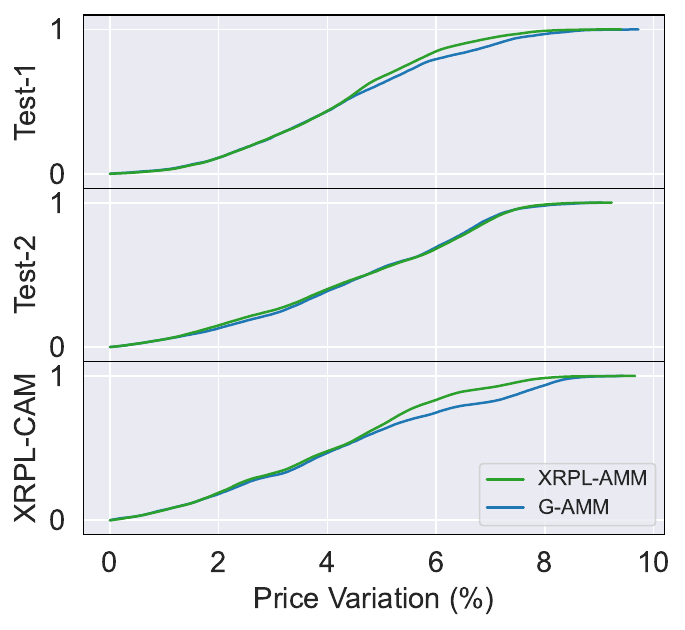}
        \captionsetup{font=scriptsize}
        \caption{Price Sync. with Reference Market for \ac{xrpl}-\ac{amm}-\ac{dex} (with and without \ac{cam}) vs. \ac{g-amm} (\ac{cdf}) using historical price data from Binance.}
        \label{fig:xrpl_vs_uniswap_price_sync_binance}
    \end{minipage}%
    \hfill%
    \begin{minipage}[t]{0.28\textwidth}
        \centering
        \includegraphics[width=\linewidth]{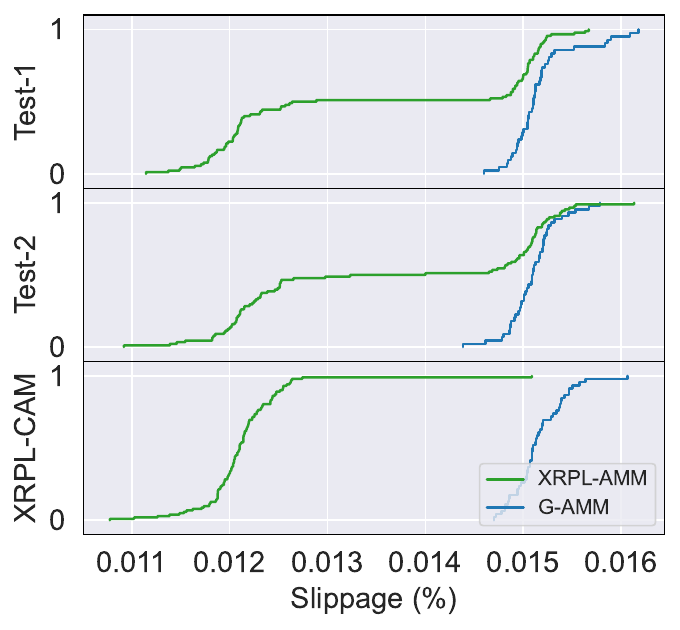}
        \captionsetup{font=scriptsize}
        \caption{Slippage for \ac{xrpl}-\ac{amm}-\ac{dex} (with and without \ac{cam}) vs \ac{g-amm} (\ac{cdf}) using historical price data from Binance.}
        \label{fig:xrpl_vs_uniswap_slippages_binance}
    \end{minipage}%
    \hfill%
    \begin{minipage}[t]{0.28\textwidth}
        \centering
        \includegraphics[width=\linewidth]{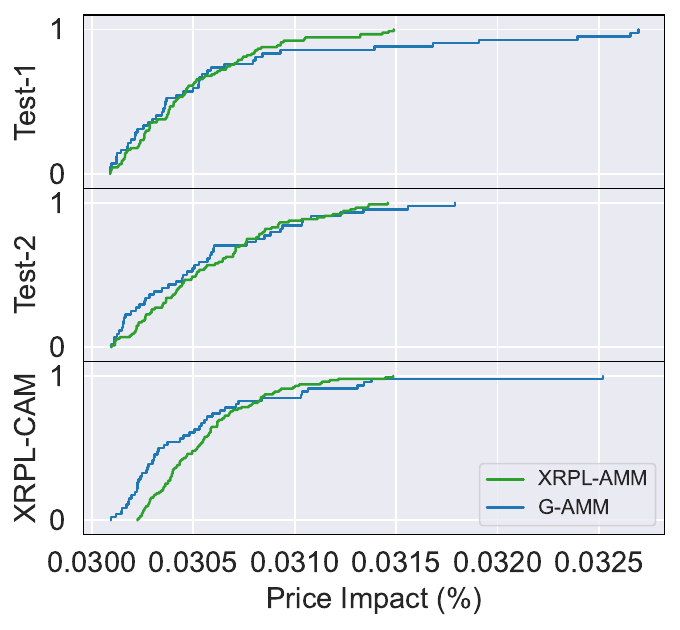}
        \captionsetup{font=scriptsize}
        \caption{Price Impact for \ac{xrpl}-\ac{amm}-\ac{dex} (with and without \ac{cam}) vs. \ac{g-amm} (\ac{cdf}) using historical price data from Binance.}
        \label{fig:xrpl_vs_uniswap_price_impact_cdf_binance}
    \end{minipage}%
    \hfill%
\end{figure*}

\subsubsection{Arbitrageurs' Profits, Transaction Cost \& Transaction Frequency}

\begin{table}[tb]
\centering
\tiny %
\captionsetup{font=scriptsize}
\caption[Test-1 and Test-2 average arbitrageurs' profits, transaction costs \& transaction frequency for \ac{xrpl}-\ac{amm}-\ac{dex} vs. \ac{g-amm}]{Test-1 and Test-2 average arbitrageurs' profits, transaction costs \& transaction frequency for \ac{xrpl}-\ac{amm}-\ac{dex} vs. \ac{g-amm}.}
\begin{tabular}{cccccc}
\toprule
\multicolumn{2}{c}{\multirow{2}{*}{\textbf{Test}}} & \multirow{2}{*}{\textbf{Profits (\ac{usdc})}} & \multirow{2}{*}{\textbf{Fees (\ac{usdc})}} & \multicolumn{2}{c}{\textbf{Transaction Count}} \\
\cmidrule{5-6}
\multicolumn{2}{c}{} & & & \textbf{Realized (\%)} & \textbf{Unrealized} \\
\midrule
\multirow{2}{*}{Test-1}
& \ac{xrpl}-\ac{amm}-\ac{dex} & 384,410 & 29 & 29 (10\%) & 260 \\
& \ac{g-amm} & 382,696 & 28 & 28 (4\%) & 634 \\
\midrule
\multirow{2}{*}{Test-2}
& \ac{xrpl}-\ac{amm}-\ac{dex} & 376,272 & 0.0003 & 27 (5\%) & 476 \\
& \ac{g-amm} & 360,581 & 107.6 & 27 (6\%) & 454 \\
\bottomrule
\end{tabular}
\label{tab:xrpl_uniswap_arb_profits_fees_txs}
\end{table}

\begin{table}[tb]
\centering
\tiny %
\captionsetup{font=scriptsize}
\caption[Average Arbitrageurs' Profits, Transaction Costs \& Transaction Count for \ac{xrpl}-\ac{cam} vs. \ac{g-amm} with different volatilities]{Average arbitrageurs' profits, transaction costs \& transaction frequency for \ac{xrpl}-\ac{cam} vs. \ac{g-amm} with different volatilities.}
\begin{tabular}{cccccc}
\toprule
\multicolumn{2}{c}{\multirow{2}{*}{\textbf{Volatility}}} & \multirow{2}{*}{\textbf{Profits (\ac{usdc})}} & \multirow{2}{*}{\textbf{Fees (\ac{usdc})}} & \multicolumn{2}{c}{\textbf{Transaction Count}} \\
\cmidrule{5-6}
\multicolumn{2}{c}{} & & & \textbf{Realized (\%)} & \textbf{Unrealized} \\
\midrule
\multirow{3}{*}{$\sigma=5\%$}
& \ac{xrpl}-\ac{cam}-A & 97,251 & \multirow{2}{*}{0.0002} & \multirow{2}{*}{16 (31.4\%)} & \multirow{2}{*}{35} \\
& \ac{xrpl}-\ac{cam}-B & 180,303 & & & \\
& \ac{g-amm} & 174,686 & 53 & 13 (4\%) & 311 \\
\midrule
\multirow{3}{*}{$\sigma=12.5\%$}
& \ac{xrpl}-\ac{cam}-A & 235,937 & \multirow{2}{*}{0.001} & \multirow{2}{*}{72 (18.3\%)} & \multirow{2}{*}{322} \\
& \ac{xrpl}-\ac{cam}-B & 823,910 & & & \\
& \ac{g-amm} & 760,056 & 230 & 58 (4.8\%) & 1,150 \\
\midrule
\multirow{3}{*}{$\sigma=20\%$}
& \ac{xrpl}-\ac{cam}-A & 468,500 & \multirow{2}{*}{0.002} & \multirow{2}{*}{159 (15.4\%)} & \multirow{2}{*}{875} \\
& \ac{xrpl}-\ac{cam}-B & 2,159,411 & & & \\
& \ac{g-amm} & 1,985,052 & 512 & 128 (4.2\%) & 2,938 \\
\bottomrule
\end{tabular}
\label{tab:xrplCAM_uniswap_profits_fees_txCount}
\end{table}

In \ac{xrpl}-\ac{cam}-B (best-case scenario for arbitrageurs), profits exceed \ac{g-amm} $70\%$ of the time at $\sigma=5\%$, and $90\%$ at $\sigma=12.5\%$ and $\sigma=20\%$. By contrast, \ac{xrpl}-\ac{amm}-\ac{dex}-\ac{cam}-A (worst-case scenario for arbitrageurs) never outperforms \ac{g-amm} since most of the profits they could have made went to liquidity providers. \autoref{tab:xrplCAM_uniswap_profits_fees_txCount} reveals:

\begin{itemize}
\item \textit{Profits}: \ac{xrpl}-\ac{cam}-B averages 7\% higher than \ac{g-amm}, while \ac{xrpl}-\ac{cam}-A is 208\% lower. This gap narrows with increased volatility, with increased arbitrage opportunities leading to higher profits.
\item \textit{Transaction Costs}: \ac{xrpl}-\ac{cam} fees are significantly lower. At $\sigma=12.5\%$, \ac{g-amm} fees ($230$ \ac{usdc}) are 23 million percent higher than \ac{xrpl}-\ac{cam} ($0.001$ \ac{usdc}). This disparity grows with volatility, with \ac{xrpl}-\ac{cam} experiencing a slight rise in fees and \ac{g-amm} seeing a more noticeable surge.
\item \textit{Transaction Count}: Both \ac{amms} see increased transactions (realized and unrealized) with higher volatility. \ac{g-amm} typically records more unrealized transactions at all volatility levels, suggesting frequent slippage condition violations. \ac{xrpl}-\ac{cam}'s realized transaction percentage decreases with volatility, while \ac{g-amm} consistently shows a lower realization rate of transactions.
\end{itemize}

For test-1 and test-2, \ac{xrpl}-\ac{amm}-\ac{dex} arbitrageurs showed $60\%$ higher profitability across both scenarios. With equalized network fees, the profit difference was minimal ($0.45\%$) but widened to $4.4\%$ with varied fees (\autoref{tab:xrpl_uniswap_arb_profits_fees_txs}). In test-2, \ac{g-amm} arbitrageurs paid 35,866,567\% more in transaction fees for the same number of transactions placed on the \ac{xrpl}-\ac{amm}-\ac{dex}. Despite \ac{g-amm} recording 2.3 times more placed transactions, only $4\%$ were realized versus $10\%$ on \ac{xrpl}-\ac{amm}-\ac{dex}.

\subsubsection{Findings using Real Market Price Data}
\label{sec:historical_data_results}

We also replicate Test-1, Test-2, and \ac{cam} tests to capture realistic market conditions using historical Binance price data. This allows us to validate our findings under empirical price dynamics. Across divergence gains, price impact, price variation, and slippage, \ac{xrpl}-\ac{amm} outperforms \ac{g-amm} consistently, and \ac{xrpl}-\ac{amm}-\ac{cam} further narrows price deviations:

\paragraph{Divergence gains} (\autoref{fig:xrpl_vs_uniswap_divergance_binance}): Divergence gains show minimal differences between the \ac{amms}, with \ac{xrpl}-\ac{amm} maintaining a slight edge across all scenarios ($+0.555$ vs $+0.552$ in Test-1; $+0.552$ vs $+0.550$ in Test-2). Adding \ac{cam} maintains this marginal advantage ($+0.555$ vs $+0.552$).

\paragraph{Price impact} (\autoref{fig:xrpl_vs_uniswap_price_impact_cdf_binance}): The \ac{xrpl}-\ac{amm} and \ac{g-amm} demonstrate nearly identical price impact in both tests ($3.05\%$ vs $3.06\%$), and this efficiency persists with \ac{cam} implementation ($3.056\%$ vs $3.052\%$).

\paragraph{Price synchronization} (\autoref{fig:xrpl_vs_uniswap_price_sync_binance}): The \ac{xrpl}-\ac{amm} achieves better price alignment in both tests ($4.22\%$ and $4.54\%$ deviation) compared to \ac{g-amm} ($4.38\%$ and $4.59\%$), with the \ac{xrpl}-\ac{amm}-\ac{dex}'s \ac{cam} feature further reducing deviation to $4.02\%$.

\paragraph{Slippage} (\autoref{fig:xrpl_vs_uniswap_slippages_binance}): \ac{xrpl}-\ac{amm} consistently outperforms \ac{g-amm} across all tests ($1.35\%$ vs $1.52\%$ in Test-1; $1.36\%$ vs. $1.51\%$ in Test-2), with its \ac{cam} feature further reducing slippage to $1.21\%$ while \ac{g-amm} remains at $1.51\%$, representing a $20\%$ improvement.

Overall, these real-data tests confirm the earlier simulation trends. Using historical Binance data, \ac{xrpl}-\ac{amm} outperforms \ac{g-amm} in price alignment ($4.22\%$ vs $4.38\%$), slippage ($1.35\%$ vs $1.52\%)$, and divergence gains ($+0.555$ vs $+0.552$), with similar price impact ($3.05\%$ vs $3.06\%$). The \ac{cam} further improves performance, reducing price synchronization to $4.02\%$ and slippage by $20\%$.

\section{Discussion}
\label{sec:discussion}

Our comparison of \ac{xrpl}-\ac{amm}-\ac{dex} (without \ac{cam}) and \ac{g-amm} reveals that \acl{xrpl} 's faster block times \cite{Perez2020} lead to better price synchronization, higher transaction realization, reduced slippage, and lower price impact. These findings align with recent empirical and theoretical studies \cite{Fritsch2024MeasuringLiquidity,Milionis2023AutomatedFees}, highlighting the importance of shorter block confirmation times for \ac{amm}-based \ac{dexs}. 

Why does blockchain infrastructure
 matter so much for \ac{amm}-based \ac{dexs}, and why should it be a consideration for their design? Unlike market makers' active role responding to trading activity in \ac{lobs} \cite{Amihud1986AssetSpread}, \ac{amms} update prices only when trades occur against the liquidity pools of their \ac{dex}. Intuitively, a blockchain infrastructure that processes transactions faster allows \ac{amm}-based \ac{dexs} to react faster to market changes with an external market, keeping prices in sync and reducing slippage. This ripple effect even touches impermanent loss, given its relationship with price slippage \cite{Milionis2022AutomatedLoss-Versus-Rebalancing}. Our results validate this intuition.

Even with faster infrastructure, \ac{amm}-based \ac{dexs} face another issue: \ac{mev} attacks. For instance, Ethereum-based \ac{dexs} are particularly vulnerable because Ethereum's transparent mempool and miner-controlled ordering \cite{ButerinEthereum:Platform.,John2025EconomicsEthereum} create a playground for attackers. Miners can cherry-pick transactions order, sparking a high-stakes race among arbitrageurs and attackers competing for prime positions in the next block \cite{Daian2020FlashInstability,Eskandari2020SoK:Blockchain,Crapis2023OptimalResources}. By contrast, the \ac{xrpl}-\ac{amm}-\ac{dex} leverages the \acl{xrpl}'s pseudo-random transaction ordering\footnote{https://github.com/XRPLF/XRPL-Standards/discussions/34}, significantly reducing the risk of front-running attacks \cite{front-run-attacks,xrpl_frontrunning,Tumas2024TheMaker}. While this does not make it immune – sandwich attacks, for example, remain a threat \cite{Tumas2024TheMaker} – it is a substantial defensive boost that could reduce price slippage. 

Beyond speed and security, the underlying infrastructure affects \ac{amm}-based \ac{dexs} in other crucial ways. Most \ac{amm}-based \ac{dexs} run on smart contracts competing for computational resources. These smart contracts can be resource-hungry, potentially consuming more gas fees than native transactions, depending on their complexity and data payloads. This resource competition could directly impact transaction costs and execution speed. Moreover, the blockchain's fee structure is pivotal in market dynamics. \ac{xrpl}-\ac{amm}-\ac{dex}'s lower network fees boost arbitrageur profits and trading volume (\autoref{sec:results}). This increased activity helps keep prices aligned with external markets. In contrast, \ac{g-amm}'s higher fees lead to broader price impact spreads, indicating more significant trade-induced market disturbances. These fee differences highlight how infrastructure choices can significantly shape an \ac{amm}-based \ac{dex}'s market efficiency and liquidity.

The \ac{xrpl}-\ac{amm}-\ac{dex}'s \ac{cam} further enhances these advantages. For \ac{lps}, \ac{xrpl}-\ac{cam} yields higher returns and lower divergence loss in their best-case scenarios, while arbitrageurs see more profits in their best-case scenario. In typical conditions, \ac{lps} benefit from higher earnings and reduced divergence loss, especially in volatile markets. This aligns with theoretical proposals for auction mechanisms in \ac{amm}-based \ac{dexs} \cite{Adams2024Am-AMM:Maker,Canidio2023ArbitrageursResponse,josojo2022MEVResearch}. Interestingly, as volatility increases, the proportion of transactions executed by auction slot holders decreases from $86\%$ at $\sigma=5\%$ to $51\%$ at $\sigma=20\%$. This trend likely results from increased competition and slippage constraints. A fascinating insight is that when we level the playing field by equalizing transaction fees and block times, the \ac{xrpl}-\ac{amm}-\ac{dex} performs remarkably similar to \ac{g-amm}. This highlights how crucial the underlying infrastructure is in shaping \ac{amm}-based \ac{dexs} dynamics.

\section{Limitations and future work}

Our agent-based simulations use simplifying assumptions such as fixed Ethereum fees, one \ac{gbm}-generated price path, constant pool sizes, single users per auction slot, and no dynamic voting or pathfinding \cite{github}. Despite these simplifications, our results align with recent research on slippage, impermanent loss, and auction mechanisms \cite{Canidio2023ArbitrageursResponse,Milionis2023AutomatedFees,Adams2024Am-AMM:Maker,josojo2022MEVResearch}, highlighting how faster block times, lower fees, and built-in auctions improve market efficiency in \ac{amm}-based \ac{dexs}. Additionally, we used block interarrival times as a proxy for infrastructure efficiency to level the playing field in benchmarking the \ac{g-amm}, akin to Uniswap V2 that runs in smart contracts, versus the \ac{xrpl}-\ac{amm}-\ac{dex} implemented at the protocol-level. Future studies could integrate more intricate factors, such as dynamic fees, diverse pool sizes, and multiple concurrent auction participants. It is important to note that the \ac{xrpl}-\ac{amm}-\ac{dex} is relatively new, launching in early 2024 \cite{XRPLAMMLaunch2024} with $\$80.37$ million \ac{tvl} \cite{XRPLAmms2024}, compared to Ethereum's $\$50.06$ billion \cite{TopChainsTVL2024} at the time of this writing. As it grows, it may face unforeseen challenges \cite{Nwaokocha2024}, and real-world adoption could impact its price synchronization and liquidity differently.

\section{Conclusion}
\label{sec:conclusion}

Our findings, using simulated and real market price data, show that the \ac{xrpl}-\ac{amm}-\ac{dex} leverages two key elements to reduce impermanent loss and price slippage: the \acl{xrpl}'s shorter block times for rapid price synchronization, and its \ac{cam} feature to incentivize beneficial arbitrage during volatility. These elements benefit both arbitrageurs and \ac{lps}, enhancing overall market efficiency. This inaugural study ventures into the unexplored domain of \ac{amm}-based \ac{dexs} in the \acl{xrpl} and provides, to the best of our knowledge, the first agent-based simulation of an auction mechanism for \ac{amm}-based \ac{dexs}, experimentally validating implications from theoretical proposals \cite{Adams2024Am-AMM:Maker,Canidio2023ArbitrageursResponse,josojo2022MEVResearch}. However, as the \ac{xrpl}-\ac{amm}-\ac{dex} is still in its early stages, at the time of this writing, its long-term success will depend on adoption rates, real-world market conditions, and its adaptability to the evolving \ac{defi} landscape.

\section*{Acknowledgements}

This work was supported by the \ac{ukcbt} and by Ripple under the \ac{ubri}~\cite{Feng2022}.

\bibliographystyle{IEEEtran}
\bibliography{references}

\end{document}